\documentclass[english,twocolumn,prb,showpacs]{revtex4}
\usepackage[T1]{fontenc}
\usepackage[latin1]{inputenc}
\usepackage{amsmath}
\usepackage{graphicx}

\makeatletter

\newcommand{\noun}[1]{\textsc{#1}}
\providecommand{\tabularnewline}{\\}

\usepackage{color}

\makeatletter

\renewcommand{\vec}[1]{\mathbf{#1}}
\usepackage{psfrag}

\usepackage{pstricks}
 
\usepackage{pst-node}

\makeatother

\usepackage{babel}
\makeatother
\begin{document}

\title{Hybrid functionals within the all-electron FLAPW method: \\
 implementation and applications of PBE0}

\author{Markus Betzinger}

\email{m.betzinger@fz-juelich.de}

\affiliation{Institut für Festkörperforschung and Institute for Advanced Simulation,
Forschungszentrum Jülich and JARA, 52425 Jülich, Germany}

\author{Christoph Friedrich}

\affiliation{Institut für Festkörperforschung and Institute for Advanced Simulation,
Forschungszentrum Jülich and JARA, 52425 Jülich, Germany}

\author{Stefan Blügel}

\affiliation{Institut für Festkörperforschung and Institute for Advanced Simulation,
Forschungszentrum Jülich and JARA, 52425 Jülich, Germany}



\begin{abstract}
We present an efficient implementation of the Perdew-Burke-Ernzerhof
hybrid functional PBE0 within the full-potential linearized augmented-plane-wave
(FLAPW) method. The Hartree-Fock exchange term, which is a central
ingredient of hybrid functionals, gives rise to a computationally
expensive nonlocal potential in the one-particle Schrödinger equation.
The matrix elements of this exchange potential are calculated with
the help of an auxiliary basis that is constructed from products of
FLAPW basis functions. By representing the Coulomb interaction in
this basis the nonlocal exchange term becomes a Brillouin-zone sum
over vector-matrix-vector products. The Coulomb matrix is calculated
only once at the beginning of a self-consistent-field cycle. We show
that it can be made sparse by a suitable unitary transformation of
the auxiliary basis, which accelerates the computation of the vector-matrix-vector
products considerably. Additionally, we exploit spatial and time-reversal
symmetry to identify the nonvanishing exchange matrix elements in
advance and to restrict the $\vec{k}$ summations for the nonlocal
potential to an irreducible set of $\vec{k}$ points. Favorable convergence
of the self-consistent-field cycle is achieved by a nested density-only
and density-matrix iteration scheme. We discuss the convergence with
respect to the parameters of our numerical scheme and show results
for a variety of semiconductors and insulators, including the oxides
$\mathrm{ZnO}$, $\mathrm{EuO}$, $\mathrm{Al_{2}O_{3}}$, and $\mathrm{SrTiO_{3}}$,
where the $\mathrm{PBE0}$ \textcolor{red}{}hybrid functional improves
the band gaps and the description of localized states in comparison
with the PBE functional. Furthermore, we find that in contrast to
conventional local exchange-correlation functionals ferromagnetic
$\mathrm{EuO}$ is correctly predicted to be a semiconductor. 
\end{abstract}
\maketitle

\section{Introduction}

Within the last decades density-functional theory (DFT) (Refs.~\onlinecite{Hohenberg-Kohn}
and \onlinecite{DFT-review}) has evolved into the state of the art
of electronic-structure calculations. It is usually applied within
the Kohn-Sham (KS) formalism,\citep{Kohn-Sham} which maps the interacting
many-electron system onto a noninteracting system with the same density.
All exchange and correlations effects of the many-electron system
are incorporated into the so-called exchange-correlation (xc) energy
functional, which is not known exactly and must be approximated in
practice. The choice of the xc functional is the only practical approximation
in this otherwise exact theory and determines the precision and efficiency
of the numerical DFT calculations.

Fortunately, already the local-density approximation (LDA),\citep{Ceperley_and_Adler,VWN}
where the xc energy functional is approximated locally by that of
the homogeneous electron gas, gives reliable results for a wide range
of materials and properties. The generalized gradient approximation
(GGA) (Refs.~\onlinecite{Generalized_Gradient_Approximation_Made_Simple}
and \onlinecite{Accurate_and_simple_density_functional_for_the_electronic_exchange_energy:_Generalized_gradient_approximation})
goes beyond this approximation by incorporating also the density gradient
of the inhomogeneous system. Due to its improved accuracy the GGA
has led to many applications of DFT in quantum chemistry. However,
there are still many cases where LDA and GGA give poor results or
are even qualitatively wrong. Among these cases are the band gaps
of solids, the atomization energies, bond lengths, and adsorption
sites of molecules as well as systems with localized states such as
transition-metal oxides. During the last decade hybrid functionals,
which combine a local or semilocal xc functional with nonlocal Hartree-Fock
(HF) exchange, have been shown to overcome these deficiencies to a
great extent.\citep{Density-functional_thermochemistry_III(Becke),Toward_reliable_density_functional_methods_without_adjustable_parameters:_The_pbe0_model,On_the_prediction_of_band_gaps_from_hybrid_functional_theory,MnO,MgO-NiO-CoO}
Hybrid functionals are usually applied within the generalized Kohn-Sham
(gKS) scheme,\citep{Generalized_Kohn-Sham_schemes_and_the_band_gap_problem}
where the HF exchange term leads to a nonlocal exchange potential
in the one-particle equations. The first hybrid functional, a half-and-half
mixing of the LDA functional with HF exchange, was proposed by Becke
in 1993.\citep{Half-and-half-mixing} Since then various \textit{ab
initio} and semiempirical hybrid functionals have been published.\citep{Rationale_for_mixing_exact_exchange_with_density_functional_approximations,Density-functional_thermochemistry_III(Becke),B3LYP,TPSSh}
The PBE0 functional,\cite{Rationale_for_mixing_exact_exchange_with_density_functional_approximations}
on which we focus in this paper, does not contain any empirical parameters
and is thus an \textit{ab initio} hybrid functional.

Hybrid functionals for systems with periodic boundary conditions were
first implemented in the late 1990s within a basis of Gaussian-type
functions and the pseudopotential plane-wave approach.\citep{Crystall98,PP-PW}
In 2005 Paier \textit{et al.}\citep{PAW-hybrid} developed an implementation
within the projector-augmented-wave (PAW) technique. In 2006 Novak
\textit{et al.}\citep{Exact_exchange_for_correlated_electrons} proposed
an approximate scheme within the full-potential linearized augmented-plane-wave
(FLAPW) approach. There the nonlocal exchange term is evaluated only
in individual atomic spheres and only for selected $l$ channels.
In this paper, we present an efficient numerical implementation of
hybrid functionals within the FLAPW method, which does not suffer
from these constraints. The FLAPW method provides a highly accurate
basis for all-electron calculations, with which a large variety of
materials, including open systems with low symmetry, $d$- and $f$-electron
systems as well as oxides, can be studied. It treats core and valence
electrons on an equal footing.

In the first Hartree-Fock implementation within the FLAPW method,
Massidda \textit{et al.}\citep{Hartree-Fock_LAPW_approach_to_the_electronic_properties_of_periodic_systems}
employed an algorithm that is routinely used to generate the potential
created by the electronic and nuclear charges and thus solves the
Poisson equation.\citep{Solution_of_poissons_equation} This \textit{Poisson
solver} can also be used for the nonlocal exchange potential because
its matrix representation involves formally identical six-dimensional
integrals over space. Instead of the real charge one then uses an
artificial charge formed by the product of two wave functions. Unfortunately,
although the algorithm is very fast, the \textit{Poisson solver} must
be called many times instead of just once when applied to the exchange
potential, which makes this approach computationally very expensive.

In this paper we propose an alternative approach that employs an auxiliary
basis, the so-called mixed product basis, which is constructed from
products of LAPW basis functions and consists of muffin-tin (MT) functions
and interstitial plane waves.\citep{All_electron_GW_approximation_with_the_mixed_basis_expansion_based_on_the_full_potential_LMTO_method}
This basis allows to decompose the state-dependent six-dimensional
integral into two three-dimensional and one state-independent six-dimensional
integral, which is the Coulomb matrix represented in the mixed product
basis. The Coulomb matrix is calculated once at the beginning of the
self-consistent-field cycle while only the three-dimensional integrals
must be evaluated in each iteration. In this formulation the matrix
elements of the nonlocal exchange potential are evaluated as Brillouin-zone
(BZ) sums over vector-matrix-vector products. Furthermore, by a suitable
unitary transformation, nearly all MT functions become multipole-free,
which makes the Coulomb matrix sparse and reduces the computational
effort for the vector-matrix-vector products considerably. As the
exchange interaction is small compared with the other energy terms,
we introduce a band cutoff as a convergence parameter and construct
the exchange matrix only in the reduced Hilbert space formed by the
wave functions up to this cutoff. In this way the number of matrix
elements that must be calculated explicitly is reduced. Because of
spatial and time-reversal symmetry some of the exchange matrix elements
vanish. In order to decide in advance, which of the matrix elements
will be nonzero, we employ a simple auxiliary operator, which has
the same symmetry properties as the nonlocal exchange potential. Additionally,
we use group theory to restrict the $\vec{k}$ summations for the
nonlocal exchange term to the smaller set of $\vec{k}$ points, which
are inequivalent with respect to the group of symmetry operations.

The long-range nature of the Coulomb interaction gives rise to a divergence
of the Coulomb matrix in the center of the BZ leading to a divergent
integrand in the exchange matrix elements. In a previous publication
we showed that the Coulomb matrix in the mixed product basis can be
decomposed exactly into a divergent and a nondivergent part.\citep{Friedrich}
The resulting divergent integrand is then given analytically and can
be integrated exactly while the nondivergent part is treated with
standard numerical integration techniques. We also calculate corrections
beyond the divergent $1/q^{2}$ term, which are obtained from $\vec{k}\cdot\vec{p}$
perturbation theory.

The paper is organized as follows. Section \ref{sec: Theory} gives
a brief introduction to hybrid functionals. Our implementation of
the nonlocal exchange potential is discussed in detail in Sec. \ref{sec: implementation}.
In Sec. \ref{sec: Results} we then apply the PBE0 hybrid functional
to prototype semiconductors and insulators and discuss the convergence
of our numerical scheme. Here we focus in particular on oxide materials.
Section \ref{sec:  Conclusions } gives a summary.

\section{Theory \label{sec: Theory}}

A hybrid functional $E_{\mathrm{xc}}^{\mathrm{hyb}}$ is a mixture
of a standard local or semilocal xc functional $E_{\mathrm{xc}}^{\mathrm{L}}=E_{\mathrm{x}}^{\mathrm{L}}+E_{\mathrm{c}}^{\mathrm{L}}$
with the exact nonlocal HF exchange energy $E_{\mathrm{x}}^{\mathrm{NL}}$
evaluated with KS wave functions\begin{eqnarray}
E_{\mathrm{xc}}^{\mathrm{hyb}} & = & (1-a)\cdot E_{\mathrm{x}}^{\mathrm{L}}+aE_{\mathrm{x}}^{\mathrm{NL}}+E_{\mathrm{c}}^{\mathrm{L}}\,,\end{eqnarray}
 where $a$ is a mixing parameter with $0<a<1$. The admixture of
$E_{\mathrm{x}}^{\mathrm{NL}}$ is motivated by the adiabatic connection
theorem\citep{The_surface_energy_of_a_bounded_electron_gas,Exchange_and_correlation_in_atoms_molecules_and_solids_by_the_spin-density-functional_formalism,The_exchange-correlation_energy_of_a_metallic_surface}
that provides an exact expression for the xc functional. This expression
becomes identical to the HF exchange term in the weakly interacting
limit, which shows that the nonlocal functional $E_{\mathrm{x}}^{\mathrm{NL}}$
is a substantial ingredient of the xc functional.

Sometimes a screened exchange term is used instead of $E_{\mathrm{x}}^{\mathrm{NL}}$.\citep{HSE,HSE-Test_on_40_semiconductors}
In some hybrid functionals one further decomposes $E_{\mathrm{x}}^{\mathrm{L}}$
and $E_{\mathrm{c}}^{\mathrm{L}}$ into the local-density and local-gradient
parts and mixes them differently.\citep{Density-functional_thermochemistry_III(Becke),B3LYP}
In this paper we focus on PBE0 (Ref.~\onlinecite{Rationale_for_mixing_exact_exchange_with_density_functional_approximations})
with the local functionals\citep{Generalized_Gradient_Approximation_Made_Simple}\begin{equation}
E_{\mathrm{x}}^{\mathrm{L}}=E_{\mathrm{x}}^{\mathrm{PBE}}\,,\quad E_{\mathrm{c}}^{\mathrm{L}}=E_{\mathrm{c}}^{\mathrm{PBE}}\end{equation}
 and the bare HF exchange term\begin{widetext}

\begin{eqnarray}
E_{\mathrm{x}}^{\mathrm{NL}} & = & -\frac{1}{2}\sum_{\sigma}\sum_{n,n'}^{\mathrm{occ.}}\sum_{\vec{k},\vec{q}}^{\mathrm{BZ}}\iint\frac{\varphi_{n\vec{k}}^{\sigma*}(\vec{r})\varphi_{n'\vec{q}}^{\sigma}(\vec{r})\varphi_{n'\vec{q}}^{\sigma*}(\vec{r}')\varphi_{n\vec{k}}^{\sigma}(\vec{r}')}{|\vec{r}-\vec{r}'|}d^{3}r\, d^{3}r'\label{eq: HF exchange energy}\end{eqnarray}
 \end{widetext}where the sum runs over the occupied (occ.) KS orbitals
$\varphi_{n\vec{k}}^{\sigma}$ of spin $\sigma$, band index $n$,
and Bloch vector $\vec{k}$. Here and in the following by a summation
over Bloch vectors $\vec{k}$ or $\vec{q}$ we mean an integration
over the Brillouin zone, which is sampled by a finite set of mesh
points. The mixing parameter $a=0.25$ was derived from first principles
in Ref.~\onlinecite{Rationale_for_mixing_exact_exchange_with_density_functional_approximations}.

Hybrid functionals are typically treated within the gKS (Ref.~\onlinecite{Generalized_Kohn-Sham_schemes_and_the_band_gap_problem})
leading to a noninteracting system of electrons that experience a
local as well as a nonlocal potential. The one-particle Schrödinger
equation in this scheme takes the form\begin{equation}
h(\vec{r})\varphi_{n\vec{k}}^{\sigma}(\vec{r})+a\int V_{\mathrm{x}}^{\mathrm{NL},\sigma}(\vec{r},\vec{r}')\varphi_{n\vec{k}}^{\sigma}(\vec{r}')d^{3}r=\epsilon_{n\vec{k}}^{\sigma}\varphi_{n\vec{k}}^{\sigma}(\vec{r})\label{eq: gKS equation}\end{equation}
 with the energy eigenvalues $\epsilon_{n\vec{k}}^{\sigma}$ and the
local one-particle Hamiltonian \begin{eqnarray}
h(\vec{r}) & = & -\frac{1}{2}\nabla^{2}+V_{\mathrm{eff}}(\vec{r})\,.\label{eq: local hamiltonian}\end{eqnarray}
 The effective potential $V_{\mathrm{eff}}(\vec{r})$ consists of
the external, Hartree, and xc potential defined by \begin{eqnarray}
V_{\mathrm{xc}}^{\mathrm{L},\sigma}(\vec{r}) & = & \frac{\delta}{\delta n^{\sigma}(\vec{r})}\left[(1-a)E_{\mathrm{x}}^{\mathrm{L}}+E_{\mathrm{c}}^{\mathrm{L}}\right]\,,\end{eqnarray}
 where the functional derivative is with respect to the electron spin
density. The nonlocal exchange potential derives from Eq.~\eqref{eq: HF exchange energy}
and is given by\begin{eqnarray}
V_{\mathrm{x}}^{\mathrm{NL},\sigma}(\vec{r},\vec{r}') & = & -\sum_{n}^{\mathrm{occ.}}\sum_{\vec{q}}^{\mathrm{BZ}}\frac{\varphi_{n\vec{q}}^{\sigma}(\vec{r})\varphi_{n\vec{q}}^{\sigma*}(\vec{r}')}{|\vec{r}-\vec{r}'|}\,.\label{eq: non-local potential}\end{eqnarray}

\section{Implementation \label{sec: implementation}}

As in a standard DFT approach, the one-particle Eqs.~\eqref{eq: gKS equation}
must be solved self-consistently because the effective potential is
a functional of the density. Apart from this, the HF exchange term
of the PBE0 hybrid functional gives rise to the nonlocal potential
in Eq.~\eqref{eq: gKS equation}, which depends on the density matrix
and, thus, on the occupied states explicitly. This is an important
issue in reaching the self-consistent solution for both the density
and the density matrix. Already in DFT calculations with local or
semilocal functionals the iterative procedure does normally not converge,
if one uses the complete \textit{output density} as input for the
next iteration. In general, one must apply a mixing scheme for the
density (or potential), e.g.,~simple or Broyden mixing,\cite{Broyden1,Broyden2}
which produces an average density out of the densities of previous
iterations as the \textit{input density} for the next step. With the
additional complication of the nonlocal exchange potential, also the
density matrix must be mixed in a suitable way, in principle. In fact,
if we simply use the \textit{output density matrix} for the next iteration,
the whole procedure takes prohibitively many steps to reach self-consistency.
We will show in Sec.~\ref{sec: Results} that with a simple trick
the number of iterations can be reduced to that of a normal DFT calculation
without the need for an explicit mixing of the density matrix.

The evaluation of the matrix elements of Eq.~\eqref{eq: non-local potential}
is by far the most time-consuming step in DFT calculations with hybrid
xc functionals. In fact, it takes much longer than any other step
in the numerical self-consistent-field cycle, even longer than the
diagonalization of the Hamiltonian matrix, which is the most time-consuming
part in calculations with local and semilocal functionals. The reason
for this is the nonlocality of the operator in Eq.~\eqref{eq: non-local potential},
which gives rise to six-dimensional integrals\begin{eqnarray}
\lefteqn{V_{\mathrm{x},nn'}^{\mathrm{NL},\sigma}(\vec{k})}\label{eq: x-potential(wf)}\\
 & = & \hspace{-0.2cm}-\sum_{n''}^{\mathrm{occ.}}\sum_{\vec{q}}^{\mathrm{BZ}}\hspace{-0.1cm}\iint\frac{\varphi_{n\vec{k}}^{\sigma*}(\vec{r})\varphi_{n''\vec{q}}^{\sigma}(\vec{r})\varphi_{n''\vec{q}}^{\sigma*}(\vec{r}')\varphi_{n'\vec{k}}^{\sigma}(\vec{r}')}{|\vec{r}-\vec{r}'|}d^{3}r\, d^{3}r'\,,\nonumber \end{eqnarray}
 whereas for the local operators in standard DFT calculations only
three-dimensional integrals must be evaluated.

In Eq.~\eqref{eq: x-potential(wf)} we have represented the exchange
operator in terms of the wave functions rather than the LAPW basis
functions. This is advantageous for two reasons: (1) if the states
$n$ and $n'$ fall into different irreducible symmetry representations,
the corresponding matrix element is zero and need not be calculated
at all (see Sec.~\ref{sub: Symmetry}). (2) Although very important,
the exchange energy is a relatively small energy contribution compared
with kinetic and potential energies. Therefore, we can afford to describe
the nonlocal exchange potential in a subspace of wave functions up
to a band cutoff $n_{\mathrm{max}}$. We only construct the matrix
for the elements with $n,n'\le n_{\mathrm{max}}$, where $n_{\mathrm{max}}$
is a convergence parameter and the rest is set to zero. We will show
in Sec.~\ref{sec: Results} that the results converge reasonably
fast with respect to $n_{\mathrm{max}}$. This reduces the computational
demand considerably.

The sum over the occupied states in Eq.~\eqref{eq: x-potential(wf)}
involves core and valence states. Core states are dispersionless,
which can be shown to lead to particularly simple and computationally
cheap expressions for their contribution to the exchange term.\citep{Core-valence-exchange}
The valence states, on the other hand, show a distinct $\vec{k}$
dependence that must be taken properly into account. Here we employ
the mixed product basis (MPB) (Refs.~\onlinecite{All_electron_GW_approximation_with_the_mixed_basis_expansion_based_on_the_full_potential_LMTO_method}
and \onlinecite{Friedrich}) that is constructed from products of
LAPW basis functions. When applied to Eq.~\eqref{eq: x-potential(wf)}
the six-dimensional integral decomposes into a vector-matrix-vector
product, where the matrix and the two vectors are the MPB representations
of the Coulomb interaction and the two wave-function products, respectively.
The Coulomb matrix is state-independent and must only be calculated
once at the beginning of the self-consistent-field cycle.

In the following we describe the implementation of the nonlocal exchange
term in detail. Sections~\ref{sec: FLAPW-method} and \ref{sub:Construction-of-the}
introduce the LAPW basis for the wave functions and the auxiliary
MPB for their products, respectively. In Sec.~\ref{sub: Sparsity}
we will show that the Coulomb matrix can be made sparse, which considerably
accelerates the vector-matrix-vector multiplications. Furthermore,
spatial and time-reversal symmetries are exploited to reduce the computational
demand, too, as we will show in Sec.~\ref{sub: Symmetry}. The Coulomb
matrix diverges in the center of the BZ. This divergence gives an
important contribution to the exchange matrix elements and must be
treated with care to guarantee a favorable convergence with respect
to the $\vec{k}$-point sampling. Sec.~\ref{sub:Treatment-of-the}
deals with this issue.

\subsection{FLAPW method \label{sec: FLAPW-method}}

In the all-electron FLAPW method\citep{FLAPW1,FLAPW2,FLAPW3} space
is partitioned into nonoverlapping atom-centered MT spheres and the
interstitial region. The core electrons, which are predominantly confined
to the MT spheres, are described by the fully relativistic Dirac equation.
For the valence electrons a basis is constructed from plane waves
in the interstitial region and numerical MT functions $u_{lp}^{a\sigma}(r)Y_{lm}(\hat{\vec{r}})$
inside the MT sphere of atom $a$, where $Y_{lm}(\hat{\vec{r}})$
denotes the spherical harmonics, $\hat{\vec{r}}=\vec{r}/r$ is a unit
vector and $\vec{r}$ is measured from the MT center located at $\vec{R}_{a}$.
The function $u_{l0}^{a\sigma}(r)$ is the solution of the radial
scalar-relativistic Dirac equation with the spherical average of the
spin-dependent effective potential and a suitably chosen energy parameter,
and $u_{l1}^{a\sigma}(r)$ is its energy derivative. In order to obtain
continuous basis functions over the whole space a linear combination
of the MT functions is matched at the sphere boundaries to each interstitial
plane wave in such a way that the resulting augmented plane waves
are continuous in value and first radial derivative. In a given unit
cell the spin-dependent basis functions with Bloch vector $\vec{k}$
are then given by\begin{widetext} \begin{equation}
\chi_{\mathbf{k}\vec{G}}^{\sigma}(\mathbf{r})=\left\{ \begin{array}{ll}
{\displaystyle \frac{1}{\sqrt{N}}\sum_{l=0}^{\mathrm{l_{\mathrm{max}}}}\sum_{m=-l}^{l}\sum_{p=0}^{1}}A_{lmp}^{a\sigma}(\mathbf{k},\vec{G})u_{lp}^{a\sigma}(|\mathbf{r}-\mathbf{R}_{a}|)Y_{lm}(\widehat{\vec{r}-\vec{R}_{a}}) & \textrm{if }\mathbf{r}\in\mathrm{MT}(a)\\
{\displaystyle \frac{1}{\sqrt{N\Omega}}}e^{i(\mathbf{k}+\mathbf{G})\cdot\mathbf{r}} & \textrm{if }\mathbf{r}\notin\mathrm{MT}\end{array}\right.\label{Eq:FLAPW-basis}\end{equation}
 \end{widetext}with the unit-cell volume $\Omega$, the number of
unit cells $N$, and reciprocal lattice vectors $\vec{G}$. The basis
functions are normalized over the whole space. For practical calculations
cutoff values for the reciprocal lattice vectors $|\vec{k}+\vec{G}|\le G_{\mathrm{max}}$
and the angular momentum $l\le l_{\mathrm{max}}$ are introduced.
For the description of semicore states the basis can be augmented
by additional functions, so-called local orbitals.\citep{Local_Orbitals1,Local_Orbitals2}
These are confined to the MT spheres and go to zero at the MT sphere
boundary.

In the LAPW basis {[}Eq.~\eqref{Eq:FLAPW-basis}] the differential
Eq.~\eqref{eq: gKS equation} becomes a generalized eigenvalue problem\begin{multline}
\sum_{\vec{G}'}\left[H_{\vec{G}\vec{G}'}^{\sigma}(\vec{k})+aV_{\mathrm{x},\vec{G}\vec{G}'}^{\mathrm{NL},\sigma}(\vec{k})\right]c_{\vec{G}'}^{\sigma}(n,\vec{k})\\
=\epsilon_{n\vec{k}}^{\sigma}\sum_{\vec{G}'}S_{\vec{G}\vec{G}'}^{\sigma}(\vec{k})c_{\vec{G}'}^{\sigma}(n,\vec{k})\,,\label{eq: APW Hamiltionian}\end{multline}
 where $H_{\vec{G}\vec{G}'}^{\sigma}$ and $V_{\mathrm{x},\vec{G}\vec{G}'}^{\mathrm{NL},\sigma}$
are the matrix representations of the operators in Eqs.~\eqref{eq: local hamiltonian}
and \eqref{eq: non-local potential}, respectively, and $S_{\vec{G}\vec{G}'}^{\sigma}$
denotes the overlap matrix. The matrix $V_{\mathrm{x},\vec{G}\vec{G}'}^{\mathrm{NL},\sigma}$
is obtained from Eq.~\eqref{eq: x-potential(wf)} by multiplying
from right and left with the inverse matrix of eigenvectors and its
adjoint, respectively,\begin{eqnarray}
V_{\mathrm{x},\vec{G}\vec{G}'}^{\mathrm{NL},\sigma}(\vec{k}) & = & \sum_{n,n'}\left[\sum_{\vec{G}''}S_{\vec{G}\vec{G}''}^{\sigma*}(\vec{k})c_{\vec{G}''}^{\sigma}(n,\vec{k})\right]V_{\mathrm{x},nn'}^{\mathrm{NL},\sigma}(\vec{k})\nonumber \\
 &  & \hphantom{z_{\vec{G}''}^{\sigma*}}\times\left[\sum_{\vec{G}''}c_{\vec{G}''}^{\sigma*}(n',\vec{k})S_{\vec{G}''\vec{G}'}^{\sigma}(\vec{k})\right]\,.\label{eq: APW trafo}\end{eqnarray}

\subsection{Mixed product basis \label{sub:Construction-of-the}}

Our implementation of the exchange potential relies on the state-independent
MPB, which is designed to represent wave-function products. The MPB
was already explained in detail in a previous publication.\citep{Friedrich}
We only sketch the main features here.

The MPB is constructed from products of LAPW basis functions, which
gives rise to interstitial plane waves (IPWs) \begin{eqnarray}
M_{\vec{G}}^{\vec{k}}(\vec{r}) & = & \frac{1}{\sqrt{N\Omega}}e^{i(\vec{k}+\vec{G})\cdot\vec{r}}\Theta(\vec{r})\end{eqnarray}
 in the interstitial region with the step function \begin{eqnarray}
\Theta(\vec{r}) & = & \left\{ \begin{array}{cl}
0 & ,\,\mathrm{if}\,\vec{r}\in\mathrm{MT}\\
1 & ,\,\mathrm{if}\,\vec{r}\notin\mathrm{MT}\end{array}\right.\,,\label{eq: step function}\end{eqnarray}
 and the set of functions $u_{lp}^{a\sigma}(r)u_{l'p'}^{a\sigma}(r)Y_{LM}(\hat{\vec{r}})$
with $|l-l'|\le L\le l+l'$ and $-L\le M\le L$ in the spheres. Usually,
the latter is highly linearly dependent. In order to remove these
linear dependences we employ a scheme proposed by Aryasetiawan and
Gunnarsson in Ref.~\onlinecite{Product-basis_method_for_calculating_dielectric_matrices}:
for each atom and $LM$ channel the overlap matrix of this set is
diagonalized; (nearly) linearly dependent combinations can then be
identified easily by small eigenvalues. We thus obtain a smaller but
still sufficiently flexible basis set by only retaining those eigenfunctions
whose eigenvalues exceed a given threshold value (typically $0.0001$).
Furthermore, we must add a spherically symmetric constant function
in order to isolate the divergent long-wavelength limit of the Coulomb
interaction (see Sec.~\ref{sub:Treatment-of-the}). In the case of
magnetic calculations the MPB is made spin-independent at this stage
by taking into account products of spin-up as well as spin-down radial
functions in the construction of the overlap matrices. From the resulting
MT functions $M_{aLMP}(\vec{r})=M_{aLP}(r)Y_{LM}(\hat{\vec{r}})$
we then formally construct Bloch functions \begin{equation}
M_{aLMP}^{\vec{k}}(\mathbf{r})=\frac{1}{\sqrt{N}}\sum_{\mathbf{T}}M_{aLMP}(\mathbf{r-T-R}_{a})e^{i\mathbf{k}\cdot\left(\mathbf{T}+\mathbf{R}_{a}\right)}\,,\label{eq: MT functions}\end{equation}
 where the sum runs over the lattice vectors and $M_{aLMP}(\vec{r})=0$,
if $r$ is larger than the MT radius $S_{a}$.

As in the case of the LAPW basis introduced in the last section, we
introduce cutoff values $G_{\mathrm{max}}^{\prime}$ and $L_{\mathrm{max}}$
for the IPWs

\begin{eqnarray}
|\vec{k}+\vec{G}| & \le & G_{\mathrm{max}}^{\prime}\le2G_{\mathrm{max}}\end{eqnarray}
 and the MT functions

\begin{eqnarray}
L & \le & L_{\mathrm{max}}\le2l_{\mathrm{max}}\,.\end{eqnarray}
 We will show in Sec.~\ref{sec: Results} that the cutoff values
can be chosen much smaller than twice the corresponding LAPW cutoff
values although this is the exact limit.

In contrast to the LAPW basis, the IPWs and MT functions are not matched
at the MT sphere boundaries but instead simply combined into the full
MPB $\{ M_{I}^{\vec{k}}(\vec{r})\}=\{ M_{aLMP}^{\vec{k}}(\vec{r}),M_{\vec{G}}^{\vec{k}}(\vec{r})\}$.
By construction the MT functions are orthonormal. As the IPWs and
the MT functions are defined in different regions of space, they do
not overlap. Only the IPWs overlap in a nontrivial way. Their overlap
matrix $O_{\vec{G}\vec{G}'}^{\vec{k}}$ is given by the Fourier transform
of the step function in Eq.~\eqref{eq: step function}

\begin{eqnarray}
O_{\vec{G}\vec{G}'}^{\vec{k}} & = & \Theta_{\vec{G}-\vec{G}'}.\end{eqnarray}

With the biorthogonal set $\{\tilde{M}_{I}^{\vec{k}}\}=\{ M_{aLMP}^{\vec{k}}(\vec{r}),\sum_{\vec{G}'}(O^{\vec{k}})_{\vec{G}'\vec{G}}^{-1}M_{\vec{G}'}^{\vec{k}}(\vec{r})\}$
we can further write the completeness relation as\begin{equation}
\sum_{I}|M_{I}^{\vec{k}}\rangle\langle\tilde{M}_{I}^{\vec{k}}|=\sum_{I}|\tilde{M}_{I}^{\vec{k}}\rangle\langle M_{I}^{\vec{k}}|=1\,,\label{eq: completeness relation}\end{equation}
 which is valid in the subspace spanned by the MPB. It is important
to note that the MPB is constructed in such a way that it describes
the wave-function products exactly in the basis-set limit.

\subsection{Sparsity of the Coulomb matrix\label{sub: Sparsity}}

With the help of the completeness relations in Eq.~\eqref{eq: completeness relation},
the integral in Eq.~\eqref{eq: x-potential(wf)} decomposes into
a vector-matrix-vector product\begin{eqnarray}
V_{\mathrm{x},nn'}^{\mathrm{NL},\sigma}(\vec{k}) & = & -\sum_{n''}^{\mathrm{occ.}}\sum_{\vec{q}}^{\mathrm{BZ}}\sum_{IJ}\langle\varphi_{n\vec{k}}^{\sigma}|\varphi_{n''\vec{k}-\vec{q}}^{\sigma}M_{I}^{\vec{q}}\rangle\nonumber \\
 &  & \times v_{IJ}(\vec{q})\langle M_{J}^{\vec{q}}\varphi_{n''\vec{k}-\vec{q}}^{\sigma}|\varphi_{n'\vec{k}}^{\sigma}\rangle\label{eq: vec-mat-vec}\end{eqnarray}
 with $n,n'\le n_{\mathrm{max}}$, where the vectors are the MPB representations
of the wave-function products and must be calculated in each iteration.
The Coulomb matrix

\begin{eqnarray}
v_{IJ}(\vec{q}) & = & \iint\frac{\tilde{M}_{I}^{\vec{q}*}(\vec{r})\tilde{M}_{J}^{\vec{q}}(\vec{r}')}{|\vec{r}-\vec{r}'|}d^{3}r\, d^{3}r'\,,\label{eq: coulomb matrix}\end{eqnarray}
 on the other hand, is independent of the wave functions and must
be constructed only once at the beginning of the self-consistent-field
cycle. It consists of four distinct blocks, the diagonal parts MT-MT
and IPW-IPW as well as the two off-diagonal parts MT-IPW and IPW-MT,
which are the complex conjugates of each other. The evaluation of
the different blocks was discussed in detail in a previous publication.\citep{Friedrich}
We choose an equidistant $\vec{k}$-point mesh in order to ensure
that $\vec{k}-\vec{q}$ is again a member of the set. Additionally,
it contains the $\Gamma$ point, which is required for a proper treatment
of the divergence of the Coulomb matrix around $\vec{q}=\vec{0}$
(see Sec.~\ref{sub:Treatment-of-the}).

\begin{figure}
\includegraphics[scale=0.5]{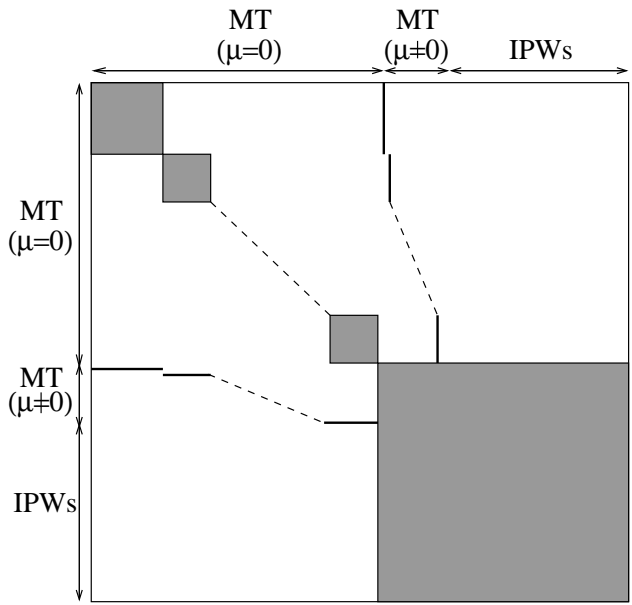}

\caption{\label{fig: coulomb matrix sparse}Illustration of the Coulomb matrix
after transforming the MPB as described in the text. The elements
that are in general nonzero are marked. The matrix is predominantly
block-diagonal: for each atom and $LM$ channel there is one block
of multipole-free MT functions $(\mu=0)$ and a larger block for the
combined set of IPWs and MT functions with a nonvanishing multipole
moment $(\mu\ne0).$ Additionally, there are very few off-diagonal
elements between MT functions with $\mu=0$ and $\mu\ne0$.}
\end{figure}

The vector-matrix-vector products must be evaluated in every iteration
for each combination of band indices $n,n'$, and $n''$ as well as
Bloch vectors $\vec{k}$ and $\vec{q}$. This easily amounts to billion
matrix operations or more and constitutes the computationally most
expensive step in the algorithm. The operations would become considerably
faster, if the Coulomb matrix could be made sparse. This is, in fact,
possible with a simple unitary transformation of the MT functions
within the subspaces of each atom and $LM$ channel. Let us consider
two MT functions with the radial parts $M_{aL1}(r)$ and $M_{aL2}(r)$.
Their electrostatic multipole moments are given by\begin{equation}
\mu_{aLP}=\int_{0}^{S_{a}}M_{aLP}(r)\, r^{L+2}dr\quad;\quad P=1,2\,.\end{equation}
 Now we apply the unitary transformation\begin{subequations}\begin{eqnarray}
M_{aL1}^{\prime}(r) & = & \frac{1}{\sqrt{\mu_{aL1}^{2}+\mu_{aL2}^{2}}}\\
 &  & \times\left[\mu_{aL1}M_{aL1}(r)+\mu_{aL2}M_{aL2}(r)\right]\nonumber \\
M_{aL2}^{\prime}(r) & = & \frac{1}{\sqrt{\mu_{aL1}^{2}+\mu_{aL2}^{2}}}\\
 &  & \times\left[\mu_{aL2}M_{aL1}(r)-\mu_{aL1}M_{aL2}(r)\right]\,,\nonumber \end{eqnarray}
 \end{subequations}which is such that the multipole moment of the
second function vanishes. With this procedure we can generally transform
a set of MT functions so that the resulting multipole moments vanish
for all but one function. For example, out of ten functions we would
obtain nine multipole-free functions and one with a nonvanishing multipole
moment. We denote the sets of these transformed functions by MT($\mu$$=$$0$)
and MT($\mu$$\ne$$0$), respectively. By construction, the former
does not generate a potential outside the MT spheres. This means that
Coulomb matrix elements in Eq.~\eqref{eq: coulomb matrix} involving
such a function can only be nonzero, if the other function is a MT
function residing in the same MT sphere; all other matrix elements
vanish. This leads to a very sparse, nearly block-diagonal form of
the Coulomb matrix illustrated in Fig.~\ref{fig: coulomb matrix sparse},
where we have ordered the MPB according to: MT($\mu$$=$$0$), MT($\mu$$\ne$$0$),
IPW. There are onsite blocks (one for each $LM$ channel) for the
MT($\mu$$=$$0$) part and one big block for the combined set of
the MT($\mu$$\ne$$0$) and the IPWs. There are only few off-diagonal
elements between MT($\mu$$=$$0$) and MT($\mu$$\ne$$0$) functions
at the same atom. Exploiting this sparsity in the matrix-vector products
drastically reduces the number of floating point operations and thus
the computational cost.%

\subsection{Symmetry\label{sub: Symmetry}}

Spatial and time-reversal symmetries are exploited to accelerate the
code in three ways: (1) inversion symmetry leads to real-valued quantities.
(2) If the wave functions $\varphi_{n\vec{k}}^{\sigma}$ and $\varphi_{n'\vec{k}}^{\sigma}$
in Eq.~\eqref{eq: vec-mat-vec} fall into different irreducible representations,
the corresponding exchange matrix element vanishes. This can be used
as a criterion whether an element must be calculated explicitly or
not. And (3), for each $\vec{k}$ chosen from the irreducible wedge
of the BZ the $\vec{q}$ summation in Eq.~\eqref{eq: vec-mat-vec}
is restricted to a smaller set of Bloch vectors giving rise to an
extended irreducible BZ.

In general, the Coulomb matrix in Eq.~\eqref{eq: coulomb matrix}
is Hermitian. If the system exhibits inversion symmetry and the MPB
functions fulfill the condition $f(-\vec{r})=f^{*}(\vec{r})$, it
becomes real-symmetric. Similarly, the vectors in Eq.~\eqref{eq: vec-mat-vec}
are then real instead of complex. This reduces the computational demand
in terms of both CPU time and memory considerably. However, presently
the condition only holds for the IPWs but not for the MT functions.
We thus combine the MT functions of each pair of atoms $a$ and $-a$,
which are related via inversion symmetry\begin{subequations}\label{sym MT functions}\begin{eqnarray}
\lefteqn{M_{aLMP}^{\prime\vec{k}}(\vec{r})}\label{eq: symmetrize MT functions}\\
 & = & \frac{1}{\sqrt{2}}\left[M_{aLMP}^{\vec{k}}(\vec{r})+\left(-1\right)^{L+M}M_{(-a)L(-M)P}^{\vec{k}}(\vec{r})\right]\nonumber \\
\lefteqn{M_{(-a)L(-M)P}^{\prime\vec{k}}(\vec{r})}\\
 & = & \frac{i}{\sqrt{2}}\left[M_{aLMP}^{\vec{k}}(\vec{r})-\left(-1\right)^{L+M}M_{(-a)L(-M)P}^{\vec{k}}(\vec{r})\right]\,.\nonumber \end{eqnarray}
 \end{subequations}If the atom is placed in the origin, i.e.,~the
atom indices $a$ and $-a$ correspond to the same atom, the transformation
in Eq.~\eqref{sym MT functions} only holds for the integer index
\textcolor{red}{}$M<0$, and we define\begin{eqnarray}
M_{aL0P}^{\prime\vec{k}}(\vec{r}) & = & \left\{ \begin{array}{ll}
\hphantom{ii\cdot}M_{aL0P}^{\vec{k}}(\vec{r})\,, & \mbox{if}\, L\,\,\mbox{even}\\
i\cdot M_{aL0P}^{\vec{k}}(\vec{r})\,, & \mbox{if}\, L\,\,\mbox{odd}\end{array}\right.\label{eq: symmetrize MT functions M=0}\end{eqnarray}
 for $M=0$. It is then easy to show that the transformed functions
will fulfill the condition above. We note that this symmetrization
leaves the form of the Coulomb matrix, as shown in Fig.~\ref{fig: coulomb matrix sparse},
intact.

The great orthogonality theorem of group theory demands that the matrix
elements $\langle\varphi_{n\vec{k}}^{\sigma}|A|\varphi_{n'\vec{k}}^{\sigma}\rangle$
of any operator $A$, which commutes with the symmetry operations
of the system, are zero, if the wave functions fall into different
irreducible representations. In particular, this holds for the exchange
operator in Eq.~\eqref{eq: non-local potential}. Unfortunately,
the irreducible representations are not available in our DFT code
and their evaluation in each iteration would be computationally expensive.
Instead, we exploit the fact that the great orthogonality theorem
applies to \textit{any} operator that has the full symmetry of the
system. A suitable operator is given by\begin{eqnarray}
\Theta^{\mathrm{MT}}(\vec{r}) & = & 1-\Theta(\vec{r})\,.\label{eq: MT stepfuntion}\end{eqnarray}
 The calculation of its matrix elements $\langle\varphi_{n\vec{k}}^{\sigma}|\Theta^{\mathrm{MT}}|\varphi_{n'\vec{k}}^{\sigma}\rangle$
is elementary and takes negligible CPU time. If the matrix elements
between two groups of degenerate wave functions are numerically zero,
we conclude that these two groups belong to different irreducible
representations. Then the corresponding matrix elements of the nonlocal
exchange potential in Eq.~\eqref{eq: vec-mat-vec} must be zero,
too. The question remains whether, conversely, the matrix elements
of Eq.~\eqref{eq: MT stepfuntion} are always nonzero, if the two
irreducible representations are identical. This is not fulfilled in
only two cases. First, either of the two wave functions is completely
confined to the MT sphere. This can be ruled out since we deal with
valence or conduction states. Second, the matrix elements are zero
by accident: the overlaps in the interstitial and the MT spheres exactly
cancel. However, this is extremely unlikely, verging on the impossible.
We find that the procedure provides a fast and reliable criterion
to decide in advance, which exchange matrix elements are nonzero and
must be calculated explicitly. In this way we again save computation
time.

In general, if a symmetry operation, which leaves the Hamiltonian
invariant, acts on a wave function, it generates another wave function
with the same energy. In other words, the solutions of the one-particle
equations at two different $\vec{k}$ points are equivalent, if the
$\vec{k}$ vectors are related by a symmetry operation. This can be
used to restrict the set of $\vec{k}$ points, at which the Hamiltonian
must be diagonalized, to a smaller set, whose members are not pairwise
related. This defines the so-called irreducible Brillouin zone (IBZ),
which is routinely employed in calculations with periodic boundary
conditions. In a similar way, the summation over $\vec{q}$ points
in the nonlocal exchange term can be confined, too. However, due to
the additional dependence on $\vec{k}$ and $\vec{k}$$-$$\vec{q}$,
we can only employ those symmetry operations $P_{i}^{\vec{k}}$ that
leave the given $\vec{k}$ vector invariant, i.e.,~$P_{i}^{\vec{k}}\vec{k}=\vec{k}+\vec{G}_{i}^{\vec{k}}$,
where $\vec{G}_{i}^{\vec{k}}$ is a reciprocal lattice vector. This
subset of operations $\{ P_{i}^{\vec{k}}\}$ is commonly called \textit{little
group} $\mathrm{LG(\vec{k})}$. In the same way as for the IBZ the
little group gives rise to a minimal set of inequivalent $\vec{q}$
points, which we denote by the extended IBZ $[\mathrm{EIBZ}(\vec{k})]$.
Leaving out here in the paper the description of nonsymmorphic and
time-reversal symmetries for simplicity, the exchange potential in
the LAPW basis can then be written as\begin{widetext}\begin{eqnarray}
V_{\mathrm{x,\vec{G}\vec{G}'}}^{\mathrm{NL},\sigma}(\vec{k}) & = & -\sum_{i}^{\mathrm{LG}(\vec{k})}\sum_{\vec{q}}^{\mathrm{EIBZ}(\vec{k})}\frac{1}{N_{\vec{k},\vec{q}}}\sum_{n}^{\mathrm{occ.}}\iint\frac{\chi_{\vec{k}\vec{G}}^{\sigma*}(\vec{r})[P_{i}^{\vec{k}^{{\scriptstyle -1}}}\varphi_{n\vec{k}-\vec{q}}^{\sigma}(\vec{r})][P_{i}^{\vec{k}^{{\scriptstyle -1}}}\varphi_{n\vec{k}-\vec{q}}^{\sigma*}(\vec{r}')]\chi_{\vec{k}\vec{G}'}^{\sigma}(\vec{r}')}{|\vec{r}-\vec{r}'|}d^{3}r\, d^{3}r'\nonumber \\
 & = & -\sum_{i}^{\mathrm{LG}(\vec{k})}\sum_{\vec{q}}^{\mathrm{EIBZ}(\vec{k})}\frac{1}{N_{\vec{k},\vec{q}}}\sum_{n}^{\mathrm{occ.}}\iint\frac{\chi_{\vec{k}(P_{i}^{\vec{k}}\vec{G}+\vec{G}_{i}^{\vec{k}})}^{\sigma*}(\vec{r})\varphi_{n\vec{k}-\vec{q}}^{\sigma}(\vec{r})\varphi_{n\vec{k}-\vec{q}}^{\sigma*}(\vec{r}')\chi_{\vec{k}(P_{i}^{\vec{k}}\vec{G}'+\vec{G}_{i}^{\vec{k}})}^{\sigma}(\vec{r}')}{|\vec{r}-\vec{r}'|}d^{3}r\, d^{3}r'\,,\label{eq: rotate APW}\end{eqnarray}
 \end{widetext}where $N_{\vec{k},\vec{q}}$ is the number of symmetry
operations that are members of both $\mathrm{LG(\vec{k})}$ and $\mathrm{LG(\vec{q})}$.
As a result, we can restrict the $\vec{q}$ summation to the $\mathrm{EIBZ}(\vec{k})$
and add the contribution of all other $\vec{q}$ points by transforming
the final matrix with the operations $\mathrm{LG}(\vec{k})$ and summing.
This takes very little computation time because a symmetry operation
acts as a one-to-one mapping in the space of the augmented plane waves
as indicated in Eq.~\eqref{eq: rotate APW}. Local orbitals transform
in a similar way. This is why we apply the symmetrization to $V_{\mathrm{x},\vec{G}\vec{G}'}^{\mathrm{NL},\sigma}$
instead of $V_{\mathrm{x},nn'}^{\mathrm{NL},\sigma}$, in which case
we would need the irreducible representations again. We note that
the whole formalism can be easily extended to the case of nonsymmorphic
and time-reversal symmetry operations.

In summary, we compute the nonlocal exchange potential $V_{\mathrm{x},nn'}^{\mathrm{NL},\sigma}(\vec{k})$
in the space of the wave functions, where we restrict the $\vec{q}$
summation to the $\mathrm{EIBZ}(\vec{k})$ and evaluate only those
band combinations $n$ and $n'$, which can be expected to be nonzero.
We then apply the transformation in Eq.~\eqref{eq: APW trafo} and
sum up the different matrix elements according to Eq.~\eqref{eq: rotate APW}.

\subsection{Singularity of the Coulomb matrix\label{sub:Treatment-of-the}}

Due to the long-range nature of the Coulomb interaction the matrix
$v_{IJ}(\vec{q})$, Eq.~\eqref{eq: coulomb matrix}, is singular
at $\vec{q}=0$, which leads to a divergent integrand in Eq.~\eqref{eq: vec-mat-vec}.
As the divergence is proportional to $1/q^{2}$, a three-dimensional
integration over the BZ yields a finite value. However, in a practical
calculation the $\vec{q}$ summation in Eq.~\eqref{eq: vec-mat-vec}
is not an integral but a weighted sum over the discrete BZ mesh. A
simple way to avoid the divergence is to exclude the point $\vec{q}=\vec{0}$
from the $\vec{k}$-point set. Then all terms in Eq.~\eqref{eq: vec-mat-vec}
are finite and the $\vec{q}$ sum can be evaluated easily. We find
that this leads to very poor convergence with respect to the BZ sampling
because the quantitatively important region around $\vec{q}=\vec{0}$
is not properly taken into account. Hence, it is advantageous to explicitly
treat the $\Gamma$ point and the singularity of the Coulomb matrix
there. This is possible by a decomposition of $v_{IJ}(\vec{q})$ into
a divergent and a nondivergent part\citep{Friedrich}\begin{equation}
v_{IJ}(\vec{q})=\frac{4\pi}{V}\frac{1}{q^{2}}\langle\tilde{M}_{I}^{\vec{q}}|e^{i\vec{q}\cdot\vec{r}}\rangle\langle e^{i\vec{q}\cdot\vec{r}}|\tilde{M}_{J}^{\vec{q}}\rangle+v_{IJ}^{\prime}(\vec{q})\,.\label{eq: div + nondiv part}\end{equation}
 The second term is finite. Its long-wavelength limit replaces the
matrix $v_{IJ}(\vec{0})$ in Eq.~\eqref{eq: vec-mat-vec} such that
the $\vec{q}$ summation can be performed numerically. The divergent
first term is given exactly in the limit $\vec{q}\rightarrow\vec{0}$
because the MPB contains the constant basis function explicitly (see
Sec.~\ref{sub:Construction-of-the}). Therefore, we may switch to
a representation with the plane waves $e^{i\vec{q}\cdot\vec{r}}$.
The contribution of the divergent term is then given by

\begin{eqnarray}
\left.V_{\mathrm{x,}nn'}^{\mathrm{NL},\sigma}(\vec{k})\right|_{\mathrm{div}} & = & -\frac{1}{2\pi^{2}}\left(\sum_{n''}^{\mathrm{occ.}}\int_{\mathrm{BZ}}\langle\varphi_{n\vec{k}}^{\sigma}|\varphi_{n''\vec{k}-\vec{q}}^{\sigma}e^{i\vec{q}\cdot\vec{r}}\rangle\frac{1}{q^{2}}\right.\nonumber \\
 &  & \left.\times\langle e^{i\vec{q}\cdot\vec{r}}\varphi_{n''\vec{k}-\vec{q}}^{\sigma}|\varphi_{n'\vec{k}}^{\sigma}\rangle d^{3}q-\mathrm{d.c.}\vphantom{\frac{1}{q^{2}}}\right),\label{eq: divergent part}\end{eqnarray}
 where $\mathrm{d.c.}$ denotes a double-counting correction for the
finite $\vec{q}$ points (see below). For the important region close
to $\vec{q}=\vec{0}$ we can replace $\langle\cdot|\cdot\rangle$
by $\delta_{nn''}$ and $\delta_{n'n''}$, respectively. We leave
out higher-order corrections here for simplicity and defer a refined
treatment employing $\vec{k}\cdot\vec{p}$ perturbation theory to
App.~\ref{sec: kp-perturtbation-theory}. In order to perform the
$q$ integration analytically, we replace $1/q^{2}$ by the function
\begin{eqnarray}
F(\vec{q}) & = & \sum_{\vec{G}}\frac{e^{-\beta|\vec{q}+\vec{G}|^{2}}}{|\vec{q}+\vec{G}|^{2}}\,,\label{eq: F(q)}\end{eqnarray}
 which was proposed by Massidda \textit{et al}.~in Ref.~\onlinecite{Hartree-Fock_LAPW_approach_to_the_electronic_properties_of_periodic_systems}.
The parameter $\beta>0$ ensures that the BZ integral can be extended
over the whole reciprocal space. In contrast to Ref.~\onlinecite{Hartree-Fock_LAPW_approach_to_the_electronic_properties_of_periodic_systems}
we choose this parameter as small as possible such that Eq.~\eqref{eq: F(q)}
is sufficiently close to $1/q^{2}$. Furthermore, by this choice terms
arising from the product of $1/q^{2}$ with the first-order term of
the exponential function are small and can thus be neglected. After
inserting Eq.~\eqref{eq: F(q)} in Eq.~\eqref{eq: divergent part}
we obtain \begin{eqnarray}
\left.V_{\mathrm{x,}nn'}^{\mathrm{NL},\sigma}(\vec{k})\right|_{\mathrm{div}} & = & -\delta_{nn'}f_{n\vec{k}}^{\sigma}\left(\frac{1}{2\pi^{2}}\int\frac{e^{-\beta|\vec{q}|^{2}}}{q^{2}}d^{3}q\vphantom{\frac{1}{N_{\vec{k}}\Omega}\sum_{q\ne0}}\right.\nonumber \\
 &  & \left.-\frac{1}{N_{\vec{k}}\Omega}\sum_{q\ne0}\frac{e^{-\beta|\vec{q}|^{2}}}{q^{2}}\right)\,,\label{eq: divergence}\end{eqnarray}
 where the summation over $q\ne0$ avoids double counting, $N_{\vec{k}}$
denotes the number of $\vec{k}$ points, and $f_{n\vec{k}}^{\sigma}$
is the occupation number. In order to evaluate the integral and sum
we introduce a reciprocal cutoff radius $q_{0}$ and finally obtain
\begin{eqnarray}
\left.V_{\mathrm{x,}nn'}^{\mathrm{NL},\sigma}(\vec{k})\right|_{\mathrm{div}} & = & -\delta_{nn'}f_{n\vec{k}}^{\sigma}\left(\frac{1}{\sqrt{\pi\beta}}\,\mathrm{erf}\left(\sqrt{\beta}q_{0}\right)\vphantom{\frac{1}{N_{\vec{k}}\Omega}\sum_{0<q\le q_{0}}}\right.\nonumber \\
 &  & \left.-\frac{1}{N_{\vec{k}}\Omega}\sum_{0<q\le q_{0}}\frac{e^{-\beta|\vec{q}|^{2}}}{q^{2}}\right)\,.\label{eq: div final}\end{eqnarray}
 We get rid off the convergence parameter $q_{0}$ by relating $\beta$
and $q_{0}$ by $e^{-\beta q_{0}^{2}}=\beta$. We find that $\beta=0.005$
is a good choice.

\begin{figure}
\includegraphics[scale=0.33,angle=-90]{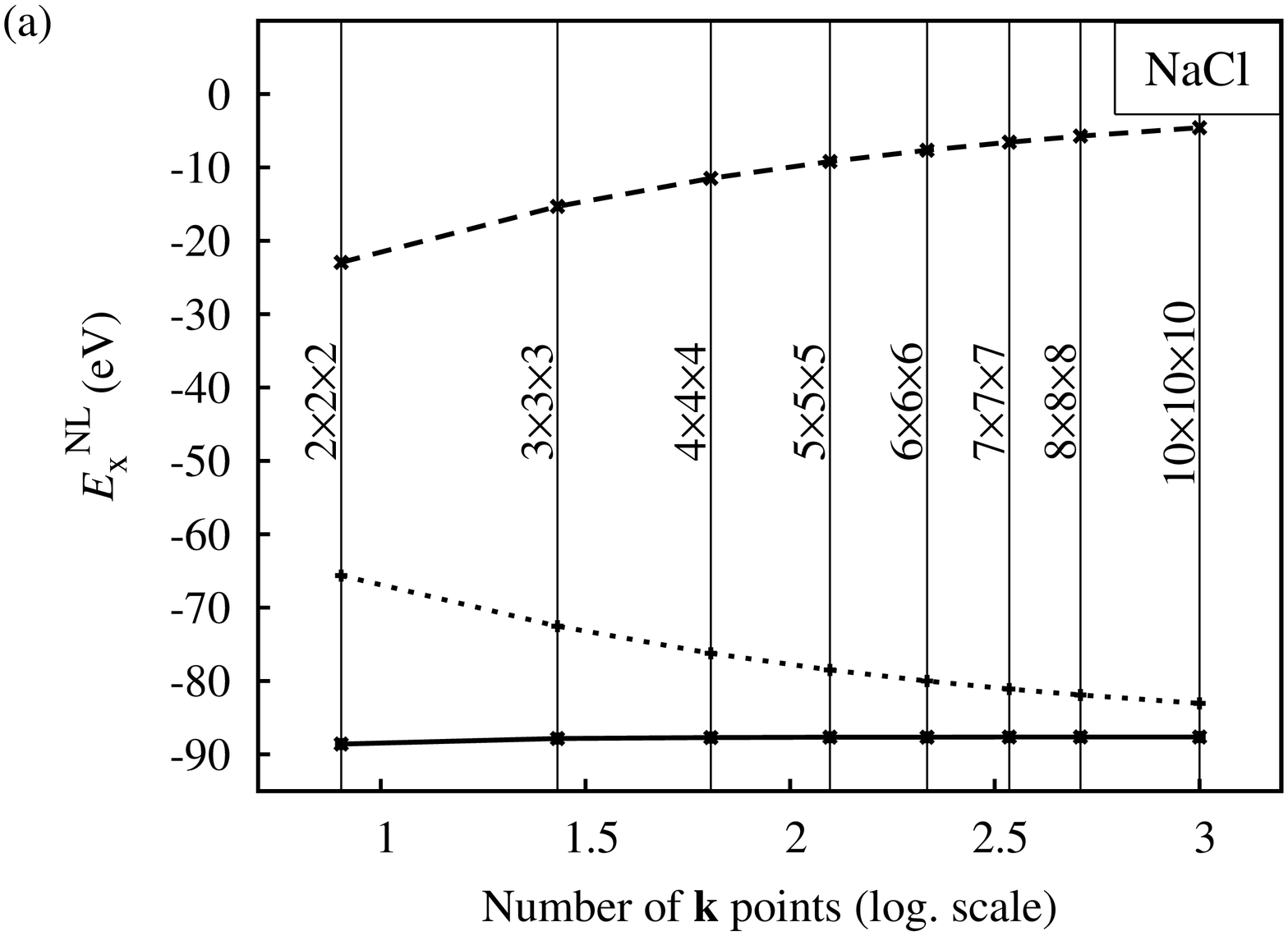}

\includegraphics[scale=0.33,angle=-90]{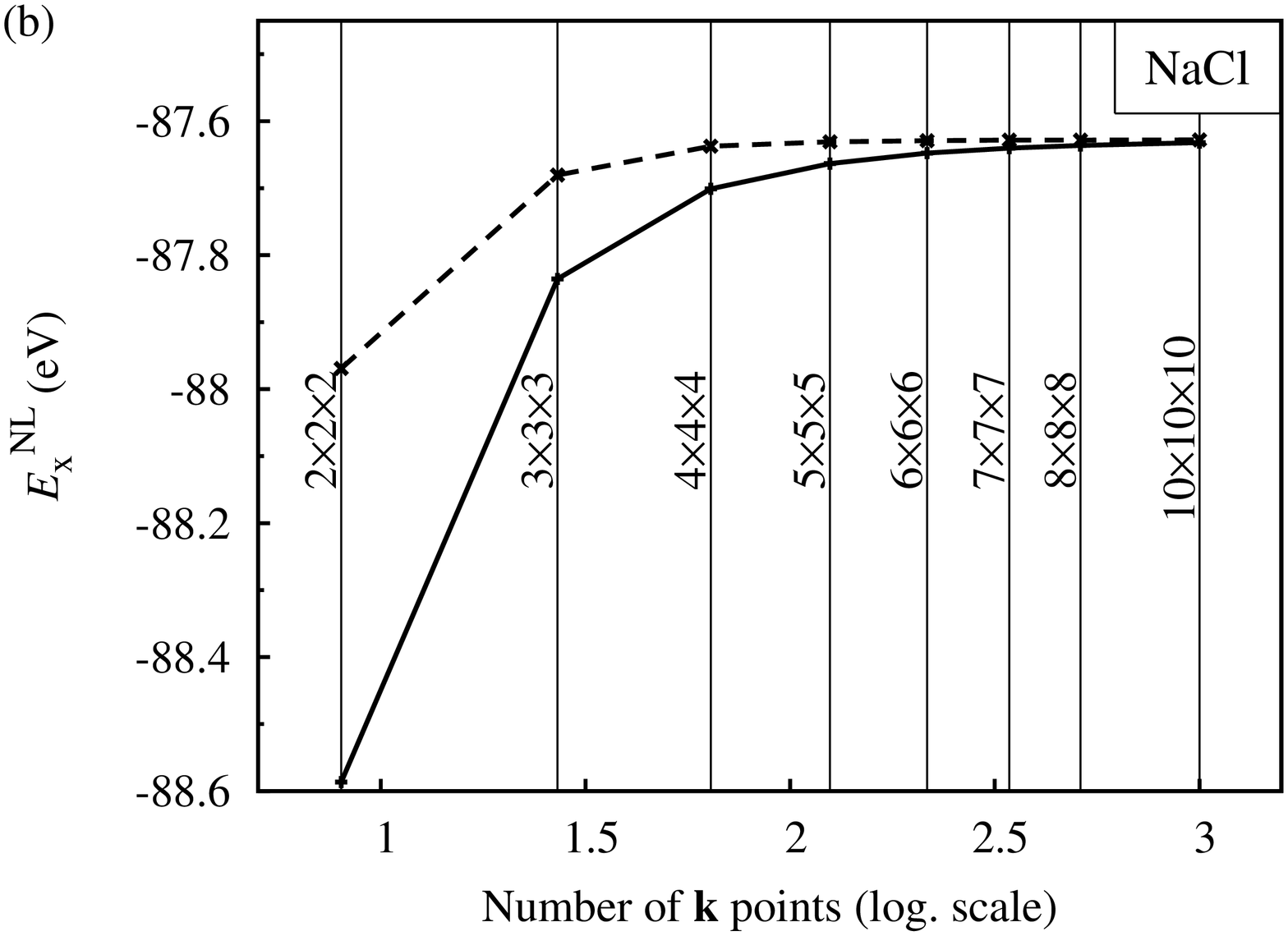}

\caption{(a) Exchange energy as a function of the $\vec{k}$-point mesh for
NaCl. The dashed and dotted curves correspond to the divergent contribution
Eq.~\eqref{eq: div final} and the remaining numerical sum, respectively.
The sum of both is shown by the solid curve. (b) Convergence of the
exchange energy with (dashed curve) and without (solid curve) higher-order
corrections at $\vec{q}=\vec{0}$. Please note the different scale
of the exchange energy in figures (a) and (b).\label{fig: k-convergence E_x }}
\end{figure}

Figure \ref{fig: k-convergence E_x } shows the convergence of the
exchange energy $E_{\mathrm{x}}^{\mathrm{NL}}=2\sum_{n\vec{k}}^{\mathrm{occ.}}V_{\mathrm{x},nn}^{\mathrm{NL}}(\vec{k})$
with respect to the $\vec{k}$-point sampling for NaCl. While the
separate contributions from the divergent term, Eq.~\eqref{eq: div final},
and the remainder converge poorly, their sum nearly look constant
on the energy scale of Fig.~\ref{fig: k-convergence E_x }(a). As
shown in Fig.~\ref{fig: k-convergence E_x }(b), the $\vec{k}$-point
convergence can be improved further by taking corrections at $\vec{q}=\vec{0}$
into account that arise from multiplying $1/q^{2}$ with second-order
terms of $\langle\cdot|\cdot\rangle\langle\cdot|\cdot\rangle$ derived
by $\vec{k}\cdot\vec{p}$ perturbation theory (see App.~\ref{sec: kp-perturtbation-theory}).%

\section{Calculations \label{sec: Results}}

We have implemented the above algorithm in the \noun{fleur} program
package.\citep{Fleur} We take the MT functions $u_{l0}^{a\sigma}(r)$
of Eq.~\eqref{Eq:FLAPW-basis} as the solutions of the radial Kohn-Sham
equation with the local effective potential derived from the full
PBE functional and $u_{l1}^{a\sigma}(r$) as their energy derivatives.
If necessary local orbitals are employed to describe semicore states.\citep{Local_Orbitals1,Local_Orbitals2}
In the calculations presented here, the core states are taken from
a preceding PBE calculation and kept fixed during the self-consistent-field
cycle with the PBE0 hybrid functional.%
\begin{figure}
\includegraphics[scale=0.33,angle=-90]{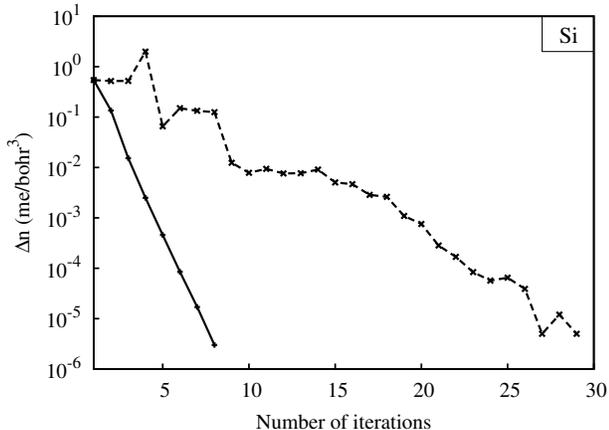}

\caption{Convergence behavior of the electron density for Si in a self-consistent-field
cycle. The solid and dashed curves correspond to calculations with
and without the nested density convergence scheme (see text).}

\label{fig: convergence speed} 
\end{figure}

The one-particle Eq.~\eqref{eq: APW Hamiltionian} must be solved
self-consistently, as its local and nonlocal potentials depend on
the electron density and density matrix, respectively, and thus on
the solution of wave functions determined in Eq.~\eqref{eq: APW Hamiltionian}.
Input and output densities must coincide in self-consistency. As a
measure of convergence one usually considers the root-mean square
of the difference between the input and output densities $\Delta n$,
measured in $\mathrm{me/bohr}^{3}$, where $\mathrm{e}$ is the elementary
charge. We consider a calculation converged, if this value falls below
$10^{-5}\,\mathrm{me/bohr}^{3}$. A straight iterative solution of
the one-particle equation is bound to diverge. In DFT calculations
with a conventional local functional it is required to construct a
new charge density for the upcoming self-consistency cycle from the
current and a history of previous densities, as for example, in the
standard simple-mixing and Broyden-mixing schemes.\cite{Broyden1,Broyden2}
However, in addition to the local effective potential Eq.~\eqref{eq: APW Hamiltionian}
contains a nonlocal potential, which depends on the density matrix
for which no similarly simple mixing procedure is available. Indeed,
we find that a standard density mixing leads to poor convergence:
$27$ iterations for $\mathrm{Si}$, as illustrated in Fig.~\ref{fig: convergence speed},
and more than $200$ iterations for $\mathrm{SrTiO}_{3}$ are necessary.
On the other hand, the fact that in the FLAPW method the basis for
the wave functions - and hence for the density matrix - depends on
the potential and changes in each iteration in contrast to that for
the density makes the definition of a mixing scheme for the density
matrix difficult, if not impossible. Therefore, we employ an alternative
pragmatic approach, which leads to a surprisingly fast density convergence
for all systems treated so far. The self-consistency cycle is divided
into an outer iteration of the density matrix and an inner self-consistency
step of the density. After the construction of the nonlocal exchange
potential we keep its matrix representation $V_{\mathrm{x},\vec{G}\vec{G}'}^{\mathrm{NL}}$
fixed and iterate Eq.~\eqref{eq: APW Hamiltionian} until self-consistency
in the density is reached; only then the exchange potential $V_{\mathrm{x},\vec{G}\vec{G}'}^{\mathrm{NL}}$
is updated from the current wave functions, which starts a new set
of inner self-consistency iterations. With this nested iterative procedure
the \textit{outer} loop converges after eight steps for $\mathrm{Si}$,
see Fig.~\ref{fig: convergence speed}, and after only twelve steps
for $\mathrm{SrTiO}_{3}$. One iteration of the \textit{inner} loop
lasts only $1.0\,\mathrm{s}$ for $\mathrm{Si}$ and $8.3\,\mathrm{s}$
for $\mathrm{SrTiO}_{3}$ on a single Intel Xeon X5355 at 2.66 GHz
(Cache 4 MB) using a 4$\times$4$\times$4 $\vec{k}$-point set. This
is negligible compared with the cost for the construction of the nonlocal
potential in the \textit{outer} loop, which takes $11.9\,\mathrm{s}$
for $\mathrm{Si}$ and $573.1\,\mathrm{s}$ for $\mathrm{SrTiO_{3}}$.

\begin{figure*}
\includegraphics[scale=0.55,angle=-90]{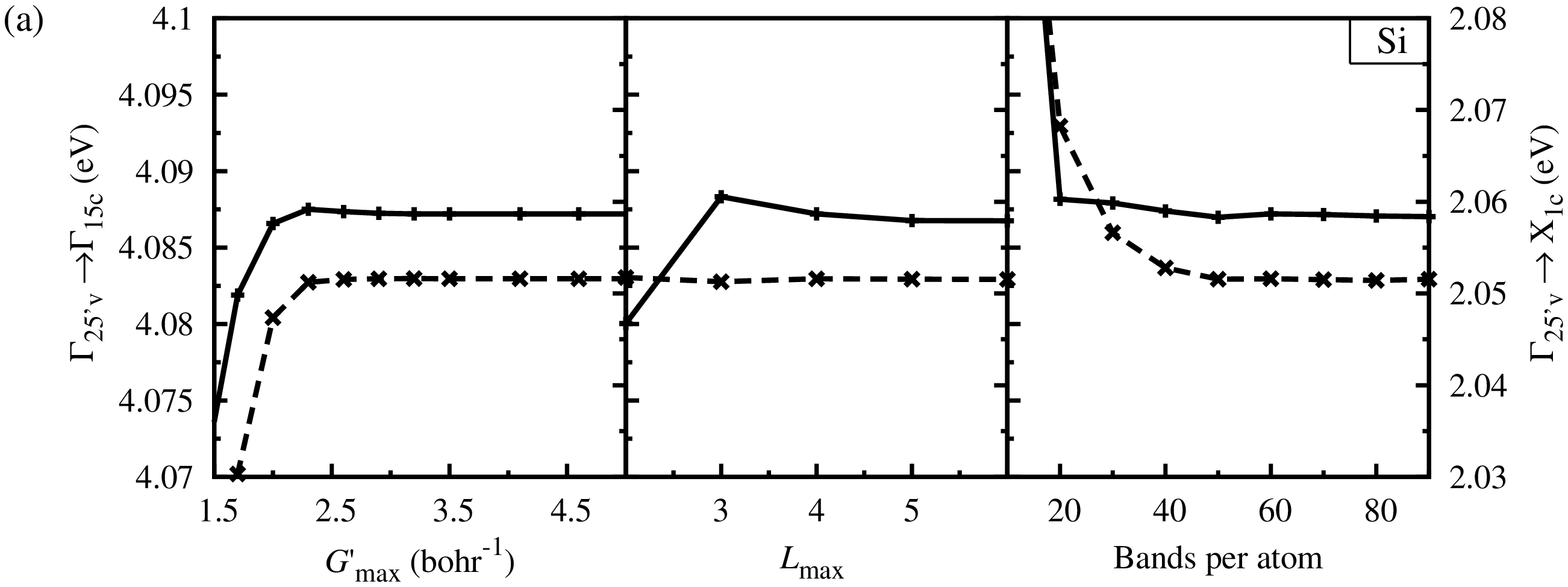}

\includegraphics[scale=0.55,angle=-90]{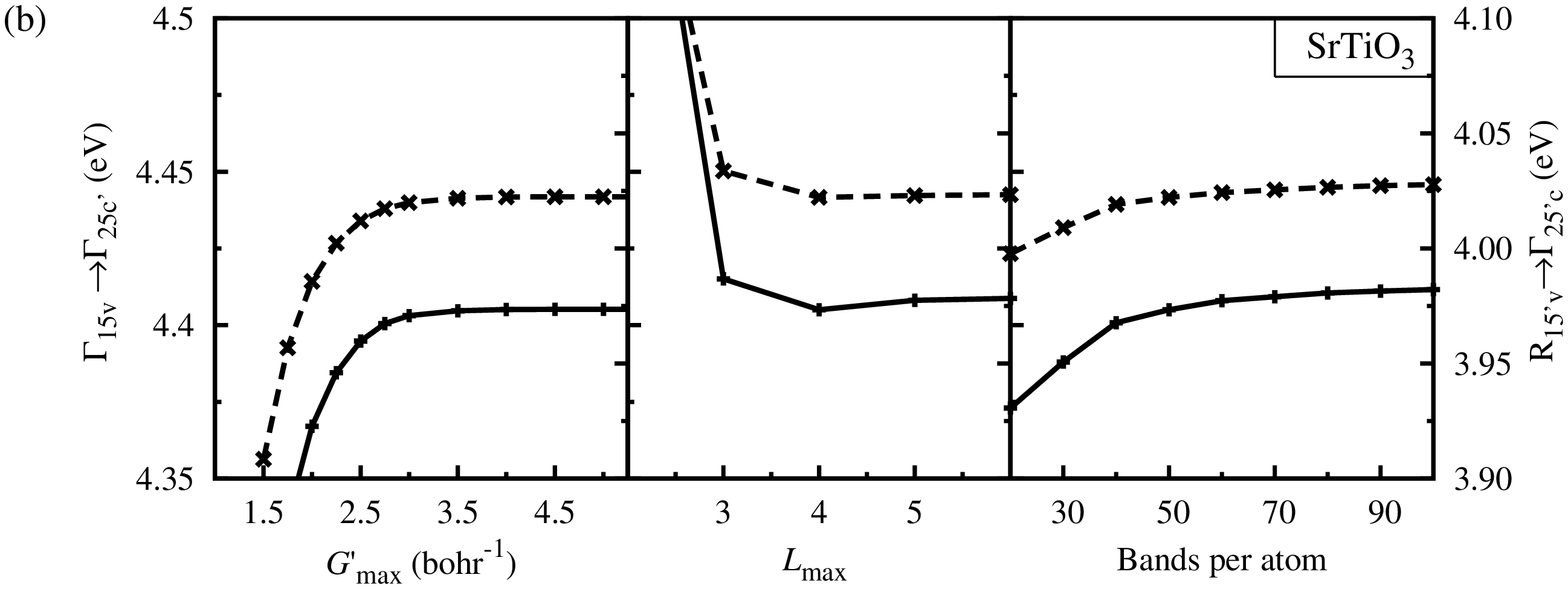}

\includegraphics[scale=0.55,angle=-90]{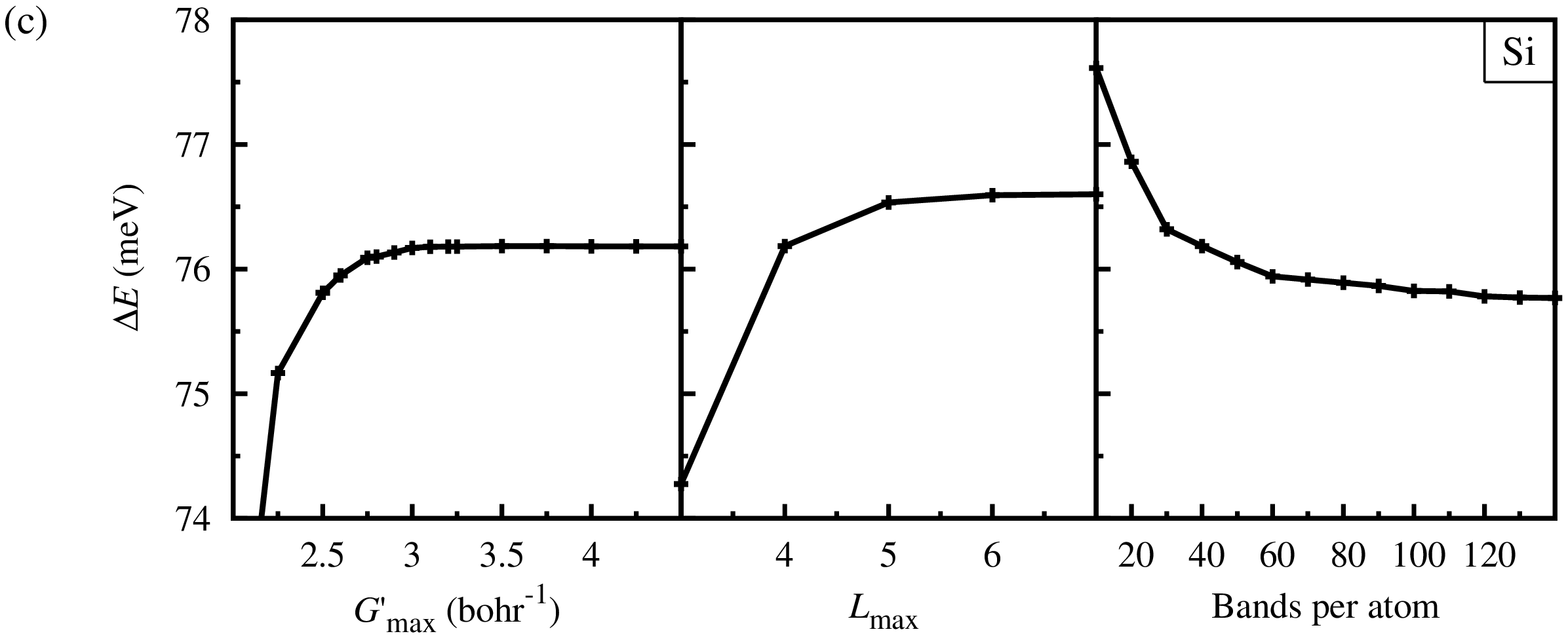}

\caption{(a) Convergence of the $\Gamma_{25'v}\rightarrow\Gamma_{15c}$ (solid
line, left scale) and $\Gamma_{25'v}\rightarrow\mathrm{X}_{1c}$ (dashed
line, right scale) transitions for $\mathrm{Si}$, (b) of the direct
(solid line, left scale) and indirect band gaps (dashed line, right
scale) of $\mathrm{SrTiO_{3}}$, and (c) the total energy difference
$\Delta E$ between the diamond (dia) and wurtzite (wz) phases of
Si ($\Delta E=E_{\mathrm{wz}}-E_{\mathrm{dia}}$) with respect to
the reciprocal cutoff value $G_{\mathrm{max}}^{\prime}$ and the angular
momentum cutoff $L_{\mathrm{max}}$ for the MPB, as well as the number
of bands per atom used to construct the exchange potential in the
space of the wave functions. For these convergence tests we employ
a 4$\times$4$\times$4 $\vec{k}$-point set. See Tables \ref{tab: FLAPW/PAW}
and \ref{tab: oxide materials} and the text for the fully converged
values. \label{fig: Si/SrTiO convergence}}
\end{figure*}

In the following we discuss the convergence of single-particle excitation
energies and total energy differences with respect to the cutoff parameters
$L_{\mathrm{max}}$ and $G'_{\mathrm{max}}$ for the MPB as well as
the number of bands $n_{\mathrm{max}}$, which are used to represent
the nonlocal exchange potential. In Figs.~\ref{fig: Si/SrTiO convergence}(a)
and \ref{fig: Si/SrTiO convergence}(b) we show the behavior of the
excitation energies of the transitions $\Gamma_{25'v}\rightarrow\Gamma_{15c}$
and $\Gamma_{25'v}\rightarrow\mathrm{X}_{1c}$ for $\mathrm{Si}$
as well as $\Gamma_{15v}\rightarrow\Gamma_{25'c}$ and $\mathrm{R}_{15'v}\rightarrow\mathrm{R}_{25'c}$
for $\mathrm{SrTiO}_{3}$ obtained from the self-consistent solution
of Eq.~\eqref{eq: APW Hamiltionian} as functions of the convergence
parameters. The diagrams show that the convergence of these transition
energies to within $0.01\,\mathrm{eV}$ is achieved for $G'_{\mathrm{max}}=2.0\,\mathrm{bohr}^{-1}$
and $G'_{\mathrm{max}}=2.7\,\mathrm{bohr}^{-1}$ for $\mathrm{Si}$
and $\mathrm{SrTiO}_{3}$, respectively. This is well below the exact
limit $G'_{\mathrm{max}}=2G_{\mathrm{max}}$ for the wave-function
products ($G_{\mathrm{max}}=3.6\,\mathrm{bohr}^{-1}$ for $\mathrm{Si}$
and $G_{\mathrm{max}}=4.3\,\mathrm{bohr}^{-1}$ for $\mathrm{SrTiO}_{3}$).
It is even below the reciprocal cutoff radius $G_{\mathrm{max}}$
for the wave functions themselves. The same can be said about the
cutoff parameter $L_{\mathrm{max}}$ for the angular momentum. Figures
\ref{fig: Si/SrTiO convergence}(a) and \ref{fig: Si/SrTiO convergence}(b)
show that for both materials $L_{\mathrm{max}}=4$ is sufficient while
the representation of wave functions that are properly matched at
the MT boundaries requires a much larger cutoff value of $l_{\mathrm{max}}=8$.
A similar behavior was found for \textit{GW} calculations employing
the MPB.\citep{GW-Fleur} The number of bands $n_{\mathrm{max}}$
that define the Hilbert space in which the exchange potential is represented
can be restricted to only $50$ bands per atom, which amounts to 100
and 250 bands for $\mathrm{Si}$ and $\mathrm{SrTiO}_{3}$, respectively.

Figure \ref{fig: Si/SrTiO convergence}(c) shows that the total energy
difference between the diamond and wurtzite structures of $\mathrm{Si}$
converges even faster than the transition energies above. With a reciprocal
cutoff radius of $G'_{\mathrm{max}}=2.25\,\mathrm{bohr^{-1}}$, an
angular-momentum cutoff of $L_{\mathrm{max}}=4$ and $20$ bands per
atom we achieve an accuracy of $1\,\mathrm{meV}$, which is one order
of magnitude smaller than the tolerance for the transition energies
and well below the error resulting from the BZ discretization of the
4$\times$4$\times$4 $\vec{k}$-point set. The calculations are converged
to within $2\,\mathrm{meV}$ with a 8$\times$8$\times$8 mesh, with
which the diamond structure is $112\,\mathrm{meV}$ lower in energy
than the wurtzite structure. This is very close to the total energy
difference of $92\,\mathrm{meV}$ obtained with the PBE functional.%

For the materials treated so far we find that $G'_{\mathrm{max}}$
can be chosen universally smaller than $G_{\mathrm{max}}$, $G'_{\mathrm{max}}=0.75\, G_{\mathrm{max}}$
as a rule of thumb, while the cutoff parameter $L_{\mathrm{max}}$
is more material specific. For example, for $\mathrm{EuO}$, whose
spin-polarized valence and conduction states are formed by $f$ electrons,
the larger value of $L_{\mathrm{max}}=6$ is necessary for proper
convergence, which is still below $l_{\mathrm{max}}=8$, though. Likewise,
the optimal number of states $n_{\mathrm{max}}$ is material specific,
and thorough convergence tests are again necessary.

For reference, we show transition energies obtained with the PBE0
hybrid functional for $\mathrm{Si}$, $\mathrm{C}$, $\mathrm{GaAs}$,
$\mathrm{MgO}$, $\mathrm{NaCl}$, and crystalline $\mathrm{Ar}$
in Table~\ref{tab: FLAPW/PAW}. All calculations are performed at
the experimental lattice constant with a 12$\times$12$\times$12
$\vec{k}$-point mesh. The PBE0 values are converged to within $0.01\,\mathrm{eV}$
with respect to the convergence parameters $G'_{\mathrm{max}}$, $L_{\mathrm{max}}$,
and $n_{\mathrm{max}}$. We also list the PBE values and a comparison
with recent PAW calculations,\citep{Screened_hybrid_density_functionals_apllied_to_solids}
with which we find an overall good agreement; for $\mathrm{Si}$ and
$\mathrm{GaAs}$ the values are nearly identical. There are slightly
larger discrepancies for systems with wider band gaps, which we attribute
to the different basis sets, because the corresponding PBE values
show similar deviations, too. In all cases the admixture of exact
HF exchange leads to an increase in the transition energies in such
a way that they come close to the measured values. For the materials
shown in Table~\ref{tab: FLAPW/PAW} there is still a slight underestimation
of the band gaps for insulators and an overestimation for semiconductors.
\begin{table}

\caption{\label{tab: FLAPW/PAW}PBE and PBE0 transition energies in eV for
Si, C, GaAs, MgO, NaCl, and Ar compared with theoretical and experimental
values from the literature. All results are obtained with a 12$\times$12$\times$12
$\vec{k}$-point set. }

\begin{ruledtabular}\begin{tabular}{ccccccc}
&
&
\multicolumn{2}{c}{This work}&
\multicolumn{2}{c}{$\mathrm{PAW^{a}}$}&
Expt.\tabularnewline
&
&
PBE&
PBE0&
PBE&
PBE0&
\tabularnewline
\hline 
Si&
$\Gamma\rightarrow\Gamma$&
2.56&
3.96&
2.57&
3.97&
$3.4^{b}$\tabularnewline
&
$\Gamma\rightarrow\mathrm{X}$&
0.71&
1.93&
0.71&
1.93&
---\tabularnewline
&
$\Gamma\rightarrow\mathrm{L}$&
1.54&
2.87&
1.54&
2.88&
$2.4^{b}$\tabularnewline
&
&
&
&
&
&
\tabularnewline
C&
$\Gamma\rightarrow\Gamma$&
5.64&
7.74&
5.59&
7.69&
$7.3^{b}$\tabularnewline
&
$\Gamma\rightarrow\mathrm{X}$&
4.79&
6.69&
4.76&
6.66&
---\tabularnewline
&
$\Gamma\rightarrow\mathrm{L}$&
8.58&
10.88&
8.46&
10.77&
---\tabularnewline
&
\multicolumn{5}{c}{}&
\tabularnewline
GaAs &
$\Gamma\rightarrow\Gamma$&
0.55&
2.02&
0.56&
2.01&
$1.63^{b}$\tabularnewline
&
$\Gamma\rightarrow\mathrm{X}$&
1.47&
2.69&
1.46&
2.67&
$2.18^{b},2.01^{b}$\tabularnewline
&
$\Gamma\rightarrow\mathrm{L}$&
1.02&
2.38&
1.02&
2.37&
$1.84^{b}$,$1.85^{b}$\tabularnewline
&
&
&
&
&
&
\tabularnewline
MgO&
$\Gamma\rightarrow\Gamma$&
4.84&
7.31&
4.75&
7.24&
$7.7^{c}$\tabularnewline
&
$\Gamma\rightarrow\mathrm{X}$&
9.15&
11.63&
9.15&
11.67&
---\tabularnewline
&
$\Gamma\rightarrow\mathrm{L}$&
8.01&
10.51&
7.91&
10.38&
---\tabularnewline
&
&
&
&
&
&
\tabularnewline
NaCl&
$\Gamma\rightarrow\Gamma$&
5.08&
7.13&
5.20&
7.26&
$8.5^{d}$\tabularnewline
&
$\Gamma\rightarrow\mathrm{X}$&
7.39&
9.59&
7.60&
9.66&
---\tabularnewline
&
$\Gamma\rightarrow\mathrm{L}$&
7.29&
9.33&
7.32&
9.41&
---\tabularnewline
&
&
&
&
&
&
\tabularnewline
Ar&
$\Gamma\rightarrow\Gamma$&
8.71&
11.15&
8.68&
11.09&
$14.2^{e}$\tabularnewline
\end{tabular}\\
 \end{ruledtabular}

\begin{raggedright}\begin{tabular}{llllll}
$^{a}$Reference \onlinecite{Screened_hybrid_density_functionals_apllied_to_solids}&
&
$^{b}$Reference \onlinecite{Landolt-Boernstein}&
&
$^{c}$Reference \onlinecite{Adachi}&
\tabularnewline
$^{d}$Reference \onlinecite{NaCl}&
&
$^{e}$Reference \onlinecite{Ar}&
&
&
\tabularnewline
\end{tabular}\par\end{raggedright}
\end{table}

\begin{figure}
\includegraphics[scale=0.9]{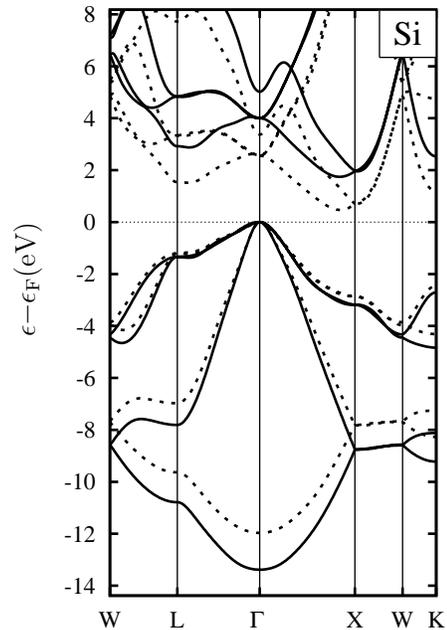}

\caption{Comparison of the PBE (dotted lines) and the Wannier-interpolated
PBE0 band structure (solid lines) for Si. Energy $\epsilon$ is given
with respect to the Fermi energy $\epsilon_{\textrm{F}}$. \label{fig: Si bandstructure}}
\end{figure}

In contrast to DFT calculations with a purely local effective potential
the nonlocality of the exchange potential does not allow a straightforward
calculation of band structures, i.e., a diagonalization of the Hamiltonian
at an arbitrary point $\vec{k}$ in the BZ because according to Eq.~\eqref{eq: vec-mat-vec}
the construction of $V_{\mathrm{x}}^{\mathrm{NL},\sigma}(\vec{k})$
would require the knowledge of all occupied states at the points $\vec{k}$$-$$\vec{q}$,
where $\vec{q}$ is an element of the $\vec{k}$ mesh defined in Sec.~\ref{sub: Sparsity}.
However, these wave functions are, in general, unknown. Therefore,
we employ the Wannier-interpolation technique\citep{Wannier,Wannier_entangled,Wannier-FLAPW}
as realized in the \noun{wannier90} code\citep{Wannier90} to interpolate
the band energies between the ones of the $\vec{k}$ mesh. As an example
Fig.~\ref{fig: Si bandstructure} shows the PBE and the interpolated
PBE0 band structure for $\mathrm{Si}$, which was constructed with
the help of eight $\mathrm{sp}{}^{3}$-like maximally localized Wannier
orbitals from the four valence and the four lowest conduction bands.
The comparison shows that the main effect of the nonlocal exchange
potential is an upwards shift of the conduction bands while the band
dispersion remains relatively unchanged. There is, however, a clear
increase in the occupied band width. As in the case of PBE the conduction-band
minimum lies at a point close to but not exactly at the X point. The
plot of the interpolated band structure thus allows to determine the
fundamental PBE0 band gap of Si easily, which amounts to $1.74\,\mathrm{eV}$.
It overestimates the experimental value of $1.17\,\mathrm{eV}$ (Ref.~\onlinecite{Landolt-Boernstein})
while the PBE value of $0.47\,\mathrm{eV}$ underestimates it. %

We finally apply the PBE0 functional to the more complex oxides $\mathrm{ZnO}$,
$\mathrm{SrTiO}_{3}$, $\alpha$-$\mathrm{Al}_{2}\mathrm{O}_{3}$,
and $\mathrm{EuO}$. The resulting band gaps are given in Table~\ref{tab: oxide materials}.%
\begin{table}

\caption{\label{tab: oxide materials}PBE and PBE0 transition energies in
eV for ZnO, $\mathrm{SrTiO}_{3}$, $\mathrm{Al}_{2}\mathrm{O}_{3}$,
and EuO. All results are obtained with a 8$\times$8$\times$8 $\vec{k}$-point
set. }

\begin{ruledtabular}

\begin{tabular}{ccccc}
&
&
PBE&
PBE0&
Expt.\tabularnewline
\hline 
ZnO &
$\Gamma\rightarrow\Gamma$&
0.94&
3.32&
$3.44^{a}$\tabularnewline
&
\multicolumn{4}{c}{}\tabularnewline
$\mathrm{SrTiO}_{3}$&
$\Gamma\rightarrow\Gamma$&
2.19&
4.39&
$3.75^{b}$\tabularnewline
&
$\mathrm{R}\rightarrow\mathrm{\Gamma}$&
1.83&
4.02&
$3.25^{b}$\tabularnewline
&
\multicolumn{4}{c}{}\tabularnewline
$\mathrm{Al}_{2}\mathrm{O}_{3}$&
$\Gamma\rightarrow\Gamma$&
6.52&
9.12&
$8.8^{c},9.5^{d}$\tabularnewline
&
\multicolumn{4}{c}{}\tabularnewline
EuO&
$\Gamma\rightarrow\mathrm{X}$&
---&
1.31&
$0.9^{e}$\tabularnewline
\end{tabular}

\end{ruledtabular}

\begin{raggedright}\begin{tabular}{llllll}
$^{a}$Reference \onlinecite{ZnO-dbands}&
&
$^{b}$Reference \onlinecite{STO-bandgap}&
&
$^{c}$Reference \onlinecite{Al2O3-2}&
\tabularnewline
$^{d}$Reference \onlinecite{Al2O3-1}&
&
$^{e}$Reference \onlinecite{EuO-FMbandgap}&
&
&
\tabularnewline
\end{tabular}\par\end{raggedright}
\end{table}

\begin{figure}
\includegraphics[scale=0.9]{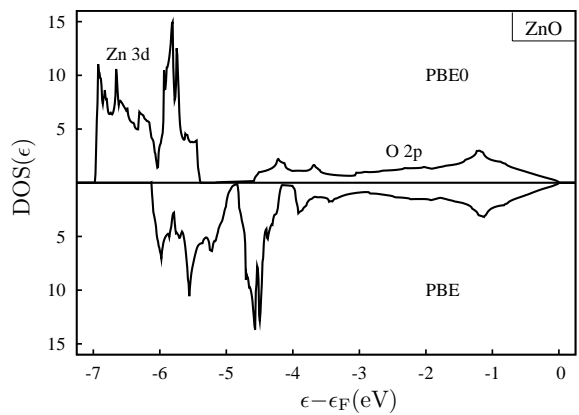}

\caption{Comparison of the PBE and PBE0 density of states (DOS) for ZnO. The
stronger binding of the Zn $d$ electrons in PBE0 is evident. Energy
$\epsilon$ is given with respect to the Fermi energy $\epsilon_{\textrm{F}}$.
\label{fig:ZnO-bandstructure}}
\end{figure}

For the II-VI semiconductor $\mathrm{ZnO}$ it is well known that
LDA and GGA not only underestimate the band gap but also yield wrong
occupied $d$ band positions;\citep{ZnO_Sic} they are about $3\,\mathrm{eV}$
too high in energy compared with experiment. As a consequence the
$\mathrm{Zn}$ $d$ states hybridize strongly with the $\mathrm{O}$
$p$ states. The wrong $d$ band position relative to the $p$ states
is commonly attributed to the unphysical self-interaction error present
in LDA and GGA, which is larger for localized than for delocalized
electrons. The admixture of exact HF exchange into the hybrid functionals
partly cancels this error and should therefore lower the relative
$d$ band position. Figure \ref{fig:ZnO-bandstructure} shows the
density of states for PBE (lower panel) and PBE0 (upper panel) again
obtained from a Wannier-interpolated denser $\vec{k}$-point mesh.
In fact, the self-interaction correction leads to a substantially
stronger binding of the Zn $d$ states from $5.1\,\mathrm{eV}$ in
PBE to $6.3\,\mathrm{eV}$ in PBE0 whereas the experimental value
is $7.8\,\mathrm{eV}$.\citep{ZnO-dbands} (The values correspond
to the center of gravity of the $d$ bands.) Concomitantly the $d$$-$$p$
hybridization becomes less pronounced leading to an increase of the
$\mathrm{O}$ $p$ valence band width from $4.2\,\mathrm{eV}$ in
PBE to $5.2\,\mathrm{eV}$ in PBE0, which deviates from the experimental
value\citep{ZnO-dbands} by only $0.1\,\mathrm{eV}$. Incidentally,
we also observe a stronger binding of the $d$ electrons in $\mathrm{Ge}$
and $\mathrm{GaAs}$. %

The missing self-interaction correction in LDA and GGA affects the
calculation of $f$-electron systems even more strongly. Ferromagnetic
$\mathrm{EuO}$ is predicted to be metallic, while experimentally
it is semiconducting with a nearly 100\% spin-polarized conduction
band,\citep{EuO-Steeneken} a property which makes $\mathrm{EuO}$
an efficient spin-filter.\citep{EuO-Moodera,EuO-Mueller} LDA+\textit{U}
has been shown to give the correct electronic structure, if the \textit{U}
parameters are chosen properly.\citep{EuO-LDA+U,EuO-Elfimov} We find
that the parameter-free PBE0 hybrid functional also gives the physically
correct semiconducting behavior for ferromagnetic EuO even though
the calculation was started from the metallic ground state obtained
from PBE. The resulting ferromagnetic semiconductor exhibits a theoretical
band gap of $1.31\,\mathrm{eV}$, which slightly overestimates the
experimental value of $0.9\,\mathrm{eV}$.\citep{EuO-FMbandgap} The
exchange splitting of the conduction band amounts to $1.02\,\mathrm{eV}$
whereas experimental measurements give $0.6\,\mathrm{eV}$.\citep{EuO-Steeneken}
In summary, the parameter-free PBE0 hybrid functional correctly predicts
the electronic ground state of ferromagnetic EuO in contrast to LDA
and GGA.

\section{Summary \label{sec:  Conclusions }}

We have presented an implementation of the PBE0 hybrid functional
within the all-electron FLAPW method as realized in the \noun{fleur}
code.\citep{Fleur} The computationally most demanding step in the
numerical procedure is the calculation of the nonlocal exact exchange
term, which is a central ingredient of hybrid functionals. Our implementation
relies on the matrix representation of the Coulomb potential in an
auxiliary mixed product basis, which is constructed from products
of the FLAPW basis functions. The nonlocal exchange integrals then
decompose into vector-matrix-vector products. The computational cost
for these products is considerably reduced by a suitable unitary transformation
of the mixed product basis, which makes the Coulomb matrix sparse.
If inversion symmetry is present, the mixed product basis can be defined
in such a way that the Coulomb matrix and the vectors become real-valued,
which again gives rise to a speedup of the code. Spatial and time-reversal
symmetries are further exploited (1) to identify those exchange matrix
elements in advance that are zero and need not to be calculated, and
(2) to restrict the $\vec{k}$-point summation for the nonlocal quantity
to an irreducible wedge of the BZ.

We have demonstrated that the PBE0 interband transition and total
energies converge quickly with respect to the parameters of the MPB.
Thus, the MPB provides a small but accurate all-electron basis for
the construction of the exchange potential. We have shown that while
a direct iteration of the generalized Kohn-Sham one-particle equation
needs extensively many steps to converge, a nested density-only and
density-matrix iteration scheme accelerates the convergence of the
self-consistent-field cycle considerably.

We confirm that the resulting PBE0 gap energies for a variety of semiconductors
and insulators are consistently closer to experimental measurements
than their PBE counterparts and compare very well with recent theoretical
results\citep{Screened_hybrid_density_functionals_apllied_to_solids}
obtained with the PAW method. In addition, we have performed PBE0
calculations for the oxides $\mathrm{ZnO}$, $\mathrm{SrTiO}_{3}$,
$\mathrm{Al}_{2}\mathrm{O}_{3}$, and $\mathrm{EuO}$. Again, the
band gaps are clearly improved compared with PBE. Here we focused
in particular on $\mathrm{ZnO}$ and $\mathrm{EuO}$, which contain
$d$ and $f$ electrons, respectively. Due to the missing self-interaction
correction conventional local xc functionals are known to fail in
describing these localized states properly: the occupied $d$ band
position of $\mathrm{ZnO}$ appear too high in energy while $\mathrm{EuO}$
is even incorrectly predicted to be a metal. The exact exchange potential
of the PBE0 hybrid functional reduces this self-interaction error
leading to an overall improved description of the relative energetic
positions of the localized and delocalized states: the Zn $d$ bands
are lowered in energy and thus come closer to their experimental position.
The reduced $d$$-$$p$ hybridization leads to an increase of the
$\mathrm{O}$ $p$ band width to $5.2\,\mathrm{eV}$, which is in
good agreement with the experiment. Furthermore, in contrast to PBE
the PBE0 hybrid functional correctly predicts a semiconducting ground
state for ferromagnetic EuO. The band gap and the energy splitting
are in satisfactory agreement with experiment.

We note that the numerical procedure presented in this paper is not
pertinent to the PBE0 hybrid functional and can easily be used for
any other hybrid functional that contains the exact exchange potential,
e.g., for the popular B3LYP functional.\citep{B3LYP} Furthermore,
it allows a straightforward implementation of more general nonlocal
potentials, e.g., the HSE functional,\citep{HSE,HSE06} which is based
on a screened Coulomb interaction. For this we just have to replace
the Coulomb matrix {[}Eq.~\ref{eq: coulomb matrix}] by the matrix
representation of the screened potential. We also note that in this
case there is no divergence at the BZ center. The orbital-dependent
Hartree-Fock term can also be used as part of an exchange-correlation
functional (e.g., the exact-exchange functional) within the optimized-effective-potential
(OEP) method.\citep{OEP1,OEP2} There, a local instead of a nonlocal
effective potential is derived, which requires the solution of the
so-called OEP equation. In general, the nonlocal exchange term is
the first term in an expansion of the xc functional with respect to
the Coulomb interaction strength and thus a central ingredient in
an systematic expansion of the xc functional.

\begin{acknowledgments}
The authors gratefully acknowledge valuable discussions with Gustav
Bihlmayer, Martin Schlipf, Frank Freimuth, Marjana Le\v{z}ai\'{c},
Yuriy Mokrousov, Tatsuya Shishidou, and Arno Schindlmayr as well as
financial support from the HGF Young Investigator Group Nanoferronics
Laboratory and the Deutsche Forschungsgemeinschaft through the Priority
Program 1145. 
\end{acknowledgments}
\appendix

\section{Terms of the BZ integrand beyond $1/q^{2}$\label{sec: kp-perturtbation-theory}}

Figure \ref{fig: k-convergence E_x }(b) shows that the $\vec{k}$-point
convergence can be improved considerably by taking into account terms
of the integrand of Eq.~\eqref{eq: divergent part} at $\vec{q}=0$
beyond the $1/q^{2}$ term

\begin{eqnarray}
A_{\vec{k},n_{1}n_{2}}^{\sigma}(\vec{q}) & = & \sum_{n}\langle\varphi_{n_{1}\vec{k}}^{\sigma}|\varphi_{n\vec{k}-\vec{q}}^{\sigma}e^{i\vec{q}\cdot\vec{r}}\rangle\frac{1}{q^{2}}\label{eq: integrand}\\
 &  & \times\langle e^{i\vec{q}\cdot\vec{r}}\varphi_{n\vec{k}-\vec{q}}^{\sigma}|\varphi_{n_{2}\vec{k}}^{\sigma}\rangle-\delta_{n_{1},n_{2}}\frac{f_{n_{1},\vec{k}}^{\sigma}}{q^{2}}\,.\nonumber \end{eqnarray}
 These terms arise from the expansion

\begin{eqnarray}
\Phi_{n,\vec{k},\vec{q}}^{\sigma}(\vec{r}) & = & e^{-i\vec{q}\cdot\vec{r}}\varphi_{n,\vec{k}+\vec{q}}^{\sigma}(\vec{r})\nonumber \\
 & = & \varphi_{n,\vec{k}}^{\sigma}(\vec{r})+\vec{q}\cdot\nabla_{\vec{q}}\Phi_{n,\vec{k},\vec{q}}^{\sigma}(\vec{r})\label{eq: kp expansion}\\
 &  & +\frac{1}{2}\vec{q}^{\mathrm{T}}\cdot\nabla_{\vec{q}}\nabla_{\vec{q}}^{\mathrm{T}}\Phi_{n,\vec{k},\vec{q}}^{\sigma}(\vec{r})\cdot\vec{q}+O(q^{3}),\nonumber \end{eqnarray}
 where \begin{equation}
\nabla_{\vec{q}}\Phi_{n,\vec{k},\vec{q}}^{\sigma}(\vec{r})=-i\sum_{n'\ne n}\frac{\langle\varphi_{n',\vec{k}}^{\sigma}|\nabla|\varphi_{n,\vec{k}}^{\sigma}\rangle}{\epsilon_{n\vec{k}}^{\sigma}-\epsilon_{n'\vec{k}}^{\sigma}}\varphi_{n',\vec{k}}^{\sigma}(\vec{r})\label{eq: first order kp expansion}\end{equation}
 is derived from $\vec{k}\cdot\vec{p}$ perturbation theory.\citep{kp-historical}
Inserting Eq.~\eqref{eq: kp expansion} into Eq.~\eqref{eq: integrand}
and using

\begin{multline}
\langle\nabla_{\vec{q}}\nabla_{\vec{q}}^{\mathrm{T}}\Phi_{n_{1},\vec{k},\vec{q}}^{\sigma}|\varphi_{n_{2},\vec{k}}^{\sigma}\rangle+\langle\varphi_{n_{1},\vec{k}}^{\sigma}|\nabla_{\vec{q}}\nabla_{\vec{q}}^{\mathrm{T}}\Phi_{n_{2},\vec{k},\vec{q}}^{\sigma}\rangle\\
=-2\langle\nabla_{\vec{q}}\Phi_{n_{1},\vec{k},\vec{q}}^{\sigma}|\nabla_{\vec{q}}^{\mathrm{T}}\Phi_{n_{2},\vec{k},\vec{q}}^{\sigma}\rangle\end{multline}
 inferred from the normalization of Eq.~\eqref{eq: kp expansion},
we obtain\begin{eqnarray}
\lefteqn{A_{\vec{k},n_{1}n_{2}}^{\sigma}(\vec{q})}\nonumber \\
 & = & -\frac{1}{q}\left(\langle\nabla_{\vec{q}}\Phi_{n_{1},\vec{k},\vec{q}}^{\sigma}|\varphi_{n_{2},\vec{k}}^{\sigma}\rangle+\langle\varphi_{n_{1},\vec{k}}^{\sigma}|\nabla_{\vec{q}}\Phi_{n_{2},\vec{k},\vec{q}}^{\sigma}\rangle\right)\hat{\vec{q}}\nonumber \\
 &  & +\hat{\vec{q}}^{\mathrm{T}}\left(\sum_{n}^{\mathrm{occ.}}\langle\varphi_{n_{1},\vec{k}}^{\sigma}|\nabla_{\vec{q}}\Phi_{n,\vec{k},\vec{q}}^{\sigma}\rangle\langle\nabla_{\vec{q}}^{\mathrm{T}}\Phi_{n,\vec{k},\vec{q}}^{\sigma}|\varphi_{n_{2},\vec{k}}^{\sigma}\rangle\right.\nonumber \\
 &  & \left.\hphantom{\vec{q}^{T}(\sum}-\langle\nabla_{\vec{q}}\Phi_{n_{1},\vec{k},\vec{q}}^{\sigma}|\nabla_{\vec{q}}^{\mathrm{T}}\Phi_{n_{2},\vec{k},\vec{q}}^{\sigma}\rangle\vphantom{\sum_{n'}^{\mathrm{occ.}}}\right)\hat{\vec{q}}\label{eq: occ.-occ.}\end{eqnarray}
 for $f_{n_{1},\vec{k}}^{\sigma}=f_{n_{2},\vec{k}}^{\sigma}=1$,\begin{eqnarray}
A_{\vec{k},n_{1}n_{2}}^{\sigma}(\vec{q}) & = & \hat{\vec{q}}^{\mathrm{T}}\left(\sum_{n}^{\mathrm{occ.}}\langle\varphi_{n_{1},\vec{k}}^{\sigma}|\nabla_{\vec{q}}\Phi_{n,\vec{k},\vec{q}}^{\sigma}\rangle\right.\nonumber \\
 &  & \left.\times\langle\nabla_{\vec{q}}^{\mathrm{T}}\Phi_{n,\vec{k},\vec{q}}^{\sigma}|\varphi_{n_{2},\vec{k}}^{\sigma}\rangle\vphantom{\sum_{n}^{\mathrm{occ.}}}\right)\hat{\vec{q}}\end{eqnarray}
 for $f_{n_{1},\vec{k}}^{\sigma}=f_{n_{2},\vec{k}}^{\sigma}=0$, and\begin{eqnarray}
\lefteqn{A_{\vec{k},n_{1}n_{2}}^{\sigma}(\vec{q})=-\frac{1}{q}\langle\nabla_{\vec{q}}\Phi_{n_{1},\vec{k},\vec{q}}^{\sigma}|\varphi_{n_{2},\vec{k}}^{\sigma}\rangle\hat{\vec{q}}}\label{eq: occ.-unocc.}\\
 &  & +\hat{\vec{q}}^{\mathrm{T}}\left(\sum_{n}^{\mathrm{occ.}}\langle\varphi_{n_{1},\vec{k}}^{\sigma}|\nabla_{\vec{q}}\Phi_{n,\vec{k},\vec{q}}^{\sigma}\rangle\times\langle\nabla_{\vec{q}}^{\mathrm{T}}\Phi_{n,\vec{k},\vec{q}}^{\sigma}|\varphi_{n_{2},\vec{k}}^{\sigma}\rangle\right)\hat{\vec{q}}\nonumber \end{eqnarray}
 for $f_{n_{1},\vec{k}}^{\sigma}=1$ and $f_{n_{2},\vec{k}}^{\sigma}=0$.
In the last case the second-order term of Eq.~\eqref{eq: kp expansion}
is neglected. The $\vec{q}$ integration in Eq.~\eqref{eq: divergent part}
finally averages over the angular-dependent terms and we are left
with\begin{widetext}\begin{eqnarray}
\overline{A}_{\vec{k},n_{1}n_{2}}^{\sigma} & = & \left\{ \begin{array}{ll}
\frac{4\pi}{3}\sum_{n}^{\mathrm{occ.}}\langle\varphi_{n_{1},\vec{k}}^{\sigma}|\nabla_{\vec{q}}^{\mathrm{T}}\Phi_{n,\vec{k}}^{\sigma}\rangle\langle\nabla_{\vec{q}}\Phi_{n,\vec{k}}^{\sigma}|\varphi_{n_{2},\vec{k}}^{\sigma}\rangle-\langle\nabla_{\vec{q}}^{\mathrm{T}}\Phi_{n_{1},\vec{k}}^{\sigma}|\nabla_{\vec{q}}\Phi_{n_{2},\vec{k}}^{\sigma}\rangle & \mathrm{for}\,\, f_{n_{1},\vec{k}}^{\sigma}=f_{n_{2},\vec{k}}^{\sigma}=1\\
\frac{4\pi}{3}\sum_{n}^{\mathrm{occ.}}\langle\varphi_{n_{1},\vec{k}}^{\sigma}|\nabla_{\vec{q}}^{\mathrm{T}}\Phi_{n,\vec{k}}^{\sigma}\rangle\langle\nabla_{\vec{q}}\Phi_{n,\vec{k}}^{\sigma}|\varphi_{n_{2},\vec{k}}^{\sigma}\rangle & \mathrm{otherwise}\end{array}\right.\,.\end{eqnarray}
 \end{widetext}These spherical averages are added to the $\vec{q}=\vec{0}$
term of the numerical integral in Eq.~\eqref{eq: divergent part}.
We note that the $1/q$ term in Eqs.~\eqref{eq: occ.-occ.} and \eqref{eq: occ.-unocc.}
exhibit an odd angular dependence and thus integrate to zero.

Recently, Shishidou and Oguchi\citep{FLAPW-kp-formalism} showed that
the fact that the LAPW basis does not fulfill\begin{equation}
e^{-i\vec{q}\cdot\vec{r}}\chi_{\vec{k}+\vec{q},\vec{G}}^{\sigma}(\vec{r})=\chi_{\vec{k},\vec{G}}^{\sigma}(\vec{r})\label{eq: non trivial k-denpedence}\end{equation}
 in the MT spheres makes a correction to Eq.~\eqref{eq: first order kp expansion}
necessary, which then becomes

\begin{eqnarray}
\lefteqn{\nabla_{\vec{q}}\Phi_{n,\vec{k}}^{\sigma}(\vec{r})=-i\sum_{n'\ne n}\frac{\langle\varphi_{n',\vec{k}}^{\sigma}|\nabla|\varphi_{n,\vec{k}}^{\sigma}\rangle}{\epsilon_{n\vec{k}}^{\sigma}-\epsilon_{n'\vec{k}}^{\sigma}}\varphi_{n',\vec{k}}^{\sigma}(\vec{r})}\\
 &  & +\left(\nabla_{\vec{q}}\tilde{\varphi}_{n,\vec{k},\vec{q}}^{\sigma}(\vec{r})-\sum_{n'}\langle\varphi_{n',\vec{k}}^{\sigma}|\nabla_{\vec{q}}\tilde{\varphi}_{n,\vec{k},\vec{q}}^{\sigma}\rangle\varphi_{n',\vec{k}}^{\sigma}(\vec{r})\right)\nonumber \end{eqnarray}

with $\tilde{\varphi}_{n,\vec{k},\vec{q}}^{\sigma}(\vec{r})=\sum_{\vec{G}}c_{\vec{G}}^{\sigma}(n,\vec{k})e^{-i\vec{q}\cdot\vec{r}}\chi_{\vec{k}+\vec{q},\vec{G}}^{\sigma}(\vec{r})$.
If Eq.~\eqref{eq: non trivial k-denpedence} was fulfilled, the correction
term would vanish because then $\tilde{\varphi}_{n,\vec{k},\vec{q}}^{\sigma}(\vec{r})=\varphi_{n,\vec{k}}^{\sigma}(\vec{r})$.
The correction only affects the last term of Eq.~\eqref{eq: occ.-occ.}
and makes the integrand exact in the limit $\vec{q}\rightarrow\vec{0}$.
However, we find that it is numerically small and negligible in the
cases treated so far.

\bibliographystyle{apsrev} \bibliographystyle{prsty}
\bibliography{biblio}

\end{document}